\documentclass[futureinternet,article,accept,pdftex,moreauthors] {Definitions/mdpi} 
\firstpage{1} 
\pubvolume{1}
\issuenum{1}
\articlenumber{0}
\pubyear{2022}
\copyrightyear{2022}
\datereceived{} 
\dateaccepted{} 
\datepublished{} 
\hreflink{https://doi.org/}

\pdfoutput=1
 
\Title{Multifractality and its sources in the digital currency market}
\TitleCitation{Multifractality sources}

\Author{Stanisław Drożdż $^{1,2,\ddagger}$\orcidA{}, Robert Kluszczyński $^{2,3,\ddagger}$\orcidB{}, Jarosław Kwapień $^{2,\ddagger}$\orcidC{} and Marcin Wątorek $^{1,\ddagger}$\orcidD{}}

\AuthorNames{Stanisław Drożdż, Robert Kluszczyński, Jarosław Kwapień and Marcin Wątorek}

\AuthorCitation{Drożdż, S.; Kluszczyński. R.; Kwapień. J.: Wątorek. M. }
\address{%
$^{1}$ \quad Faculty of Computer Science and Mathematics, Cracow University of Technology, ul.~Warszawska 24, 31-155 Krak\'ow, Poland \\
$^{2}$ \quad Complex Systems Theory Department, Institute of Nuclear Physics, Polish Academy of Sciences, Radzikowskiego 152, 31-342 Kraków, Poland \\
$^{3}$ \quad Faculty of Mathematics and Computer Science, Jagiellonian University, ul.~Łojasiewicza 6, 
30-348 Kraków, Poland \\
}
\corres{Correspondence: jaroslaw.kwapien@ifj.edu.pl (J.K.)}

\secondnote{These authors contributed equally to this work.}

\abstract{Multifractality in time series analysis characterizes the presence of multiple scaling exponents, indicating heterogeneous temporal structures and complex dynamical behaviors beyond simple monofractal models. In the context of digital currency markets, multifractal properties arise due to the interplay of long-range temporal correlations and heavy-tailed distributions of returns, reflecting intricate market microstructure and trader interactions. Incorporating multifractal analysis into the modeling of cryptocurrency price dynamics enhances the understanding of market inefficiencies, may improve volatility forecasting and facilitate the detection of critical transitions or regime shifts. Based on the multifractal cross-correlation analysis (MFCCA) whose spacial case is the multifractal detrended fluctuation analysis (MFDFA), as the most commonly used practical tools for quantifying multifractality, in the present contribution a recently proposed method of disentangling sources of multifractality in time series was applied to the most representative instruments from the digital market. They include Bitcoin (BTC), Ethereum (ETH), decentralized exchanges (DEX) and non-fungible tokens (NFT). The results indicate the significant role of heavy tails in generating a broad multifractal spectrum. However, they also clearly demonstrate that the primary source of multifractality are temporal correlations in the series, and without them, multifractality fades out. It appears characteristic that these temporal correlations, to a large extent, do not depend on the thickness of the tails of the fluctuation distribution. These observations, made here in the context of the digital currency market, provide a further strong argument for the validity of the proposed methodology of disentangling sources of multifractality in time series.
}

\keyword{complexity; time series analysis; nonlinear correlations; multifractality; singularity spectra; $q$-Gaussian distributions; digital financial instruments}

\begin{document}

\section{Introduction}

The rapid evolution of digital currency markets that has followed introduction of Bitcoin in 2009 — the first asset entirely based on the then newly introduced blockchain technology~\cite{NakamotoS-2008a} — has substantially influenced the global financial landscape~\cite{WatorekM-2021a}. As decentralized technologies challenge traditional monetary frameworks, cryptocurrencies such as Bitcoin, Ethereum, and thousands of altcoins have introduced new dynamics characterized by extreme volatility~\cite{CorbetS-2022a,EvrimMandaciP-2022a,NguyenAPN-2023a}, heavy-tailed distributions~\cite{DrozdzS-2018a,StosicD-2018a,JamesN-2021b,WatorekM-2023b}, and long-range temporal correlations~\cite{DrozdzS-2018a,TakaishiT-2018a,takaishi2020m,JamesN-2021a,james2022c,watorekfutnet2022,Brouty2024,BuiHQ-2025a}. These dynamics render conventional econometric models insufficient, necessitating more sophisticated tools to understand the intrinsic complexity of digital asset price behavior~\cite{KwapienJ-2012a,WatorekM-2021a,JamesN-2021b,ManaviSA-2020a,SpurrA-2021a,JamesN-2022a}. One such tool is multifractal analysis~\cite{JiangZQ-2019a}, a framework that extends beyond linear descriptions to capture variability in scaling properties across time and magnitudes of fluctuations.

Multifractality, or multifractal scaling, refers to the presence of multiple scaling exponents in a time series, indicating heterogeneity in local singularities~\cite{HalseyTC-1986a}. Unlike monofractal processes, which exhibit uniform scaling behavior, multifractal processes capture a spectrum of exponents that reflect the system’s complexity across scales. In financial contexts, this translates to the convoluted periods of high and low volatility~\cite{FarmerJD-2004a}, the related clustering of extreme events~\cite{MandelbrotBB-1963a}, and long-range dependences~\cite{Ausloos2002,KUTNER2004,ContR-2007a,OhG-2008a}. These features are prominently observable in digital currency markets, where trading occurs globally, continuously, and with limited regulation~\cite{WatorekM-2021a}. Yet, despite widespread recognition of multifractality in financial data, its sources remain a subject of ongoing debate~\cite{ZhouWX-2009a,BarunikJ-2012a,GreenD-2014a,MoralesR-2014a,Klamut2018,KutnerR-2019a,KwapienJ-2023a}.

A recent study by Kluszczyński \textit{et al.}~\cite{KluszczynskiR-2025a} addresses this issue on a fully quantitative level by rigorously examining the relative role of the two allegedly principal sources of multifractality: (1) temporal correlations, particularly long-range nonlinear correlations, and (2) heavy-tailed probability distribution functions (PDFs) of fluctuations. Their findings ultimately disqualify the dichotomy still appearing in the literature that treats these sources as independent contributors. Using a combination of synthetic cascades and empirical data, they demonstrate that true multifractality requires temporal correlations — heavy tails alone cannot produce multifractal spectra unless such correlations are present. In fact, fat-tailed distributions, modeled via $q$-Gaussian frameworks, were found to merely modulate the width of the multifractal spectrum, but only in the presence of underlying correlations, as already indicated by earlier published works~\cite{DrozdzS-2009a,ZhouWX-2012a,KwapienJ-2023a}. 

This insight has direct relevance to digital currency markets. Cryptocurrencies are notorious for their non-Gaussian returns, often displaying kurtosis even exceeding that of traditional financial assets~\cite{WatorekM-2021b}. However, the findings of Ref.~\cite{KluszczynskiR-2025a} imply that such statistical events do not, on their own, confirm multifractality. To discern whether the multifractal behavior in crypto markets is genuine or an artifact of finite sample size and fat tails, one must isolate and analyze the role of correlation structure. This can be done by applying multifractal detrended fluctuation analysis (MFDFA)~\cite{KantelhardtJ-2002a} as the central tool, alongside reshuffling and $q$-Gaussian filtering techniques, to distinguish between sources of multifractal complexity.

For the global digital currency markets, this methodological rigor is especially crucial. The presence of algorithmic trading~\cite{MakarovI-2020a,CohenG-2022a,FangF-2022a,CohenG-2023a,Watorek2023c}, speculative bubbles~\cite{GerlachJC-2019a,HuberTA-2022a}, and geopolitical shocks~\cite{BouriE-2022a,HongMY-2022a,KhalfaouiR-2023a,WatorekM-2023a,FangY-2024a} introduces complex dependencies that may mimic or obscure genuine multifractal behavior. Also, social media influence on investor behavior can be substantial~\cite{PoongodiM-2021a,AharonDY-2022a} with the NFT market being particularly susceptible to this effect as a single message may cause a significant increase in volatility~\cite{SzydloP-2024a,WatorekM-2024a}. By leveraging the framework developed in~\cite{KluszczynskiR-2025a}, this paper aims to systematically assess the multifractal properties of major cryptocurrencies across global exchanges. In particular, we examine how the singularity spectrum widths and asymmetries evolve with market maturity, regulation, and technological development.

Our present contribution is twofold. First, we apply advanced multifractal analysis techniques to a comprehensive dataset of global digital currency price series, including Bitcoin (BTC), Ethereum (ETH), also from the decentralized trading, and emerging non-fungible token market, spanning several years and exchanges. Second, by adopting the disentangling approach~\cite{KluszczynskiR-2025a}, we present evidence that observed multifractality is driven by temporal correlations — indicative of market memory and structure — and, when such temporal correlations related to volatility clustering are present, the distributional features such as fat tails and extreme events significantly broaden the singularity spectrum. This inspection provides deeper insight into the dynamics of digital asset markets and offers a more solid foundation for modeling them.

\section{Materials and Methods}
\label{sect::methods}

\subsection{Data and its characteristics}

In this study, we examine high-frequency data obtained from Binance~\cite{binance}, the largest cryptocurrency exchange by daily trading volume~\cite{coinmarketcap}. Specifically, we analyze the price time series $\{p(t_i)\}_{i=1}^T$ for two major cryptocurrencies: BTC and ETH, expressed in USDT. The data were sampled at regular intervals of $\Delta t = 1$ min, where $\Delta t = t_{i+1} - t_i$ and $i = 1, \ldots, T-1$. The chosen 1-minute resolution is the highest possible to minimize the number of zero log-returns (such returns usually distort results of a multifractal analysis at small time scales comparable with the average zero-return interval). Reducing the resolution by taking larger $\Delta t$ would reduce the thickness of the distribution's tails~\cite{WatorekM-2021b}.

\begin{figure}[ht!]
\centering
\includegraphics[width=0.99\textwidth]{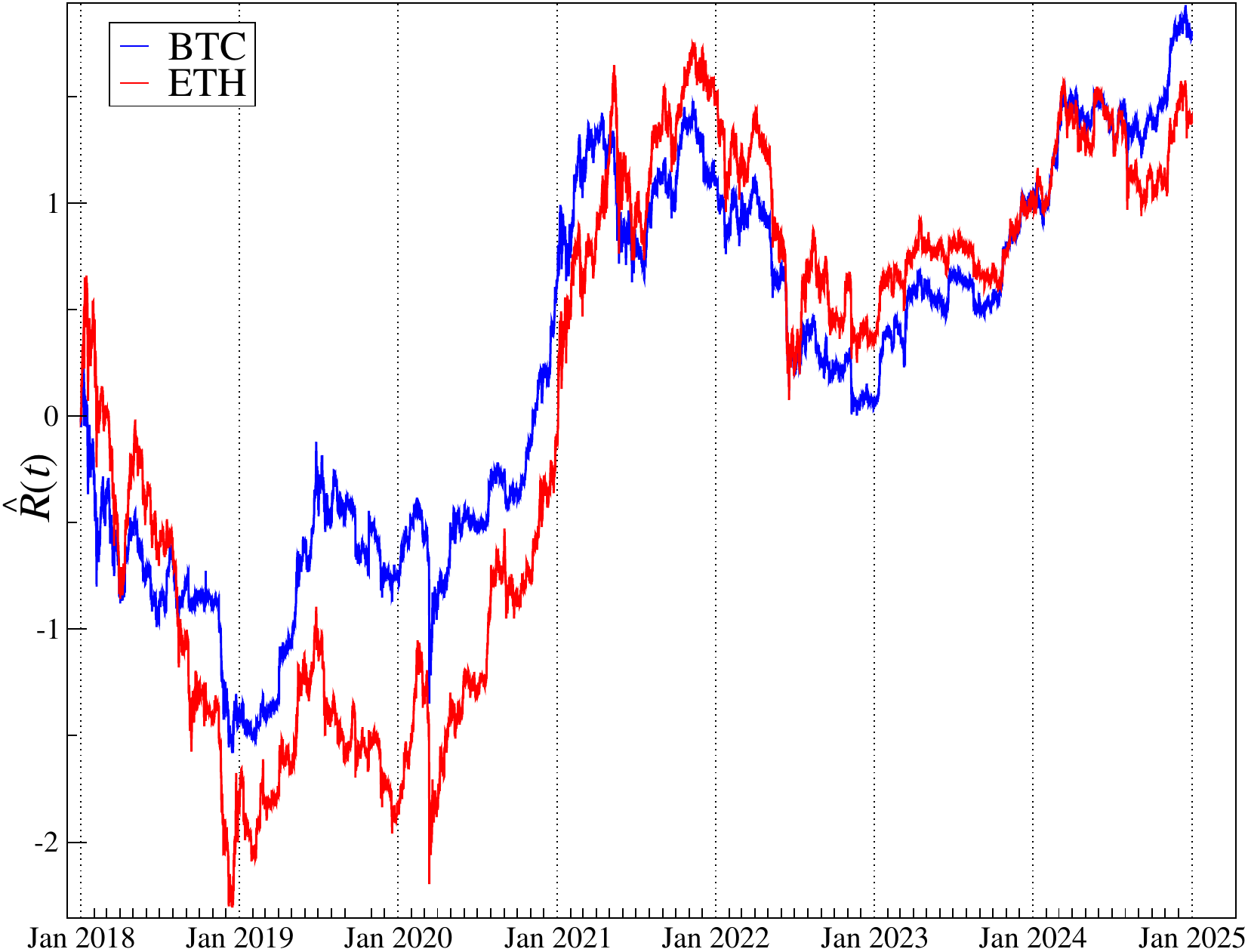}
\caption{Evolution of the cumulative log-returns $\hat{R}(t)$ of the BTC and ETH  over the time period from Jan 1, 2018 to Dec 31, 2024.}
\label{fig::returns.integrated}
\end{figure}

Given that cryptocurrency markets operate continuously, 24 hours a day, 7 days a week, our dataset spanning the period from January 1, 2018, to December 31, 2024 includes 2557 consecutive trading days with a total of $T = 3,682,080$ data points. This time frame covers the most dynamic periods — including the Covid-19 pandemic~\cite{JamesN-2021b,Crane2022,AMMYDRISS2023} — in the evolution of the two most important cryptocurrencies, i.e., BTC and ETH. From these time series, the logarithmic price returns $R(t_i) = \ln p(t_{i+1}) - \ln p(t_i)$ are calculated. In terms of the cumulative log-returns ${\hat R} (t_i) = \sum_{k=1}^{k=i} R(t_k)$, the relative price changes of these two cryptocurrencies are illustrated in Fig.(~\ref{fig::returns.integrated}). Such a representation allows us for direct tracking of relative changes in these financial instruments. In the present case of BTC and ETH, the overall trend of their price changes is seen to go largely in parallel in the period considered. 

The period before 2018 was not included in the analysis because, as shown in our previous work~\cite{DrozdzS-2018a,WatorekM-2021a}, sufficient scaling to determine the characteristics of multifractality has been being observed only since mid-2017. Moreover, the selected data range corresponds to their availability at the appropriate frequency on the Binance exchange, which became the most frequently traded only in 2018~\cite{KwapienJ-2022a}.


\begin{figure}[ht!]
\centering
\includegraphics[width=0.99\textwidth]{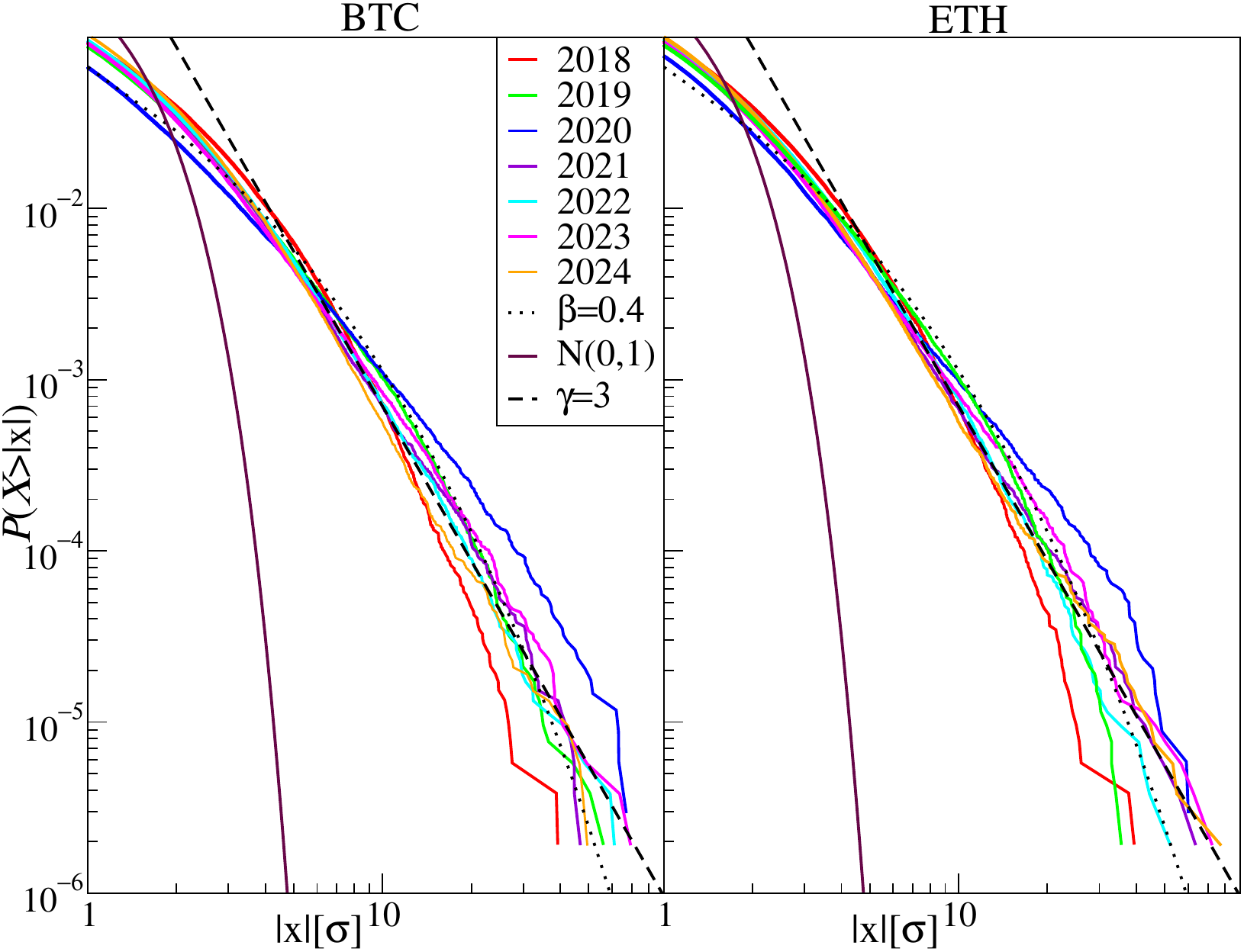}
\caption{Cumulative distribution function for BTC and ETH by year together with Gaussian, power-law (with $\gamma=3$) and stretched exponential distributions (with $\beta=0.4$).}
\label{fig::returns.pdf}
\end{figure}

The heterogeneity of the cryptocurrency market dynamics over this 7-year period is evident in the distributions of fluctuations in the corresponding 1-min return rates in subsequent 1-year periods, as shown in Fig.~\ref{fig::returns.pdf}. Of course, these distributions are always fat-tailed, but the tail thickness changes quite significantly in relation to the two reference distributions, which can here be considered, i.e., the stretched exponential $P(X>x) \sim \exp (-x^{\beta})$ and the inverse-cubic power law $P(X>x) \sim x^{-\gamma}$. As it can be seen, the thickest tails of these distributions, both for BTC and ETH, are observed in 2020, the year of greatest anxiety related to the Covid-19 pandemic. The fitting exponents of the distributions for each year are presented in Tab.~\ref{data-spec}.


\begin{figure}[ht!]
\centering
\includegraphics[width=0.99\textwidth]{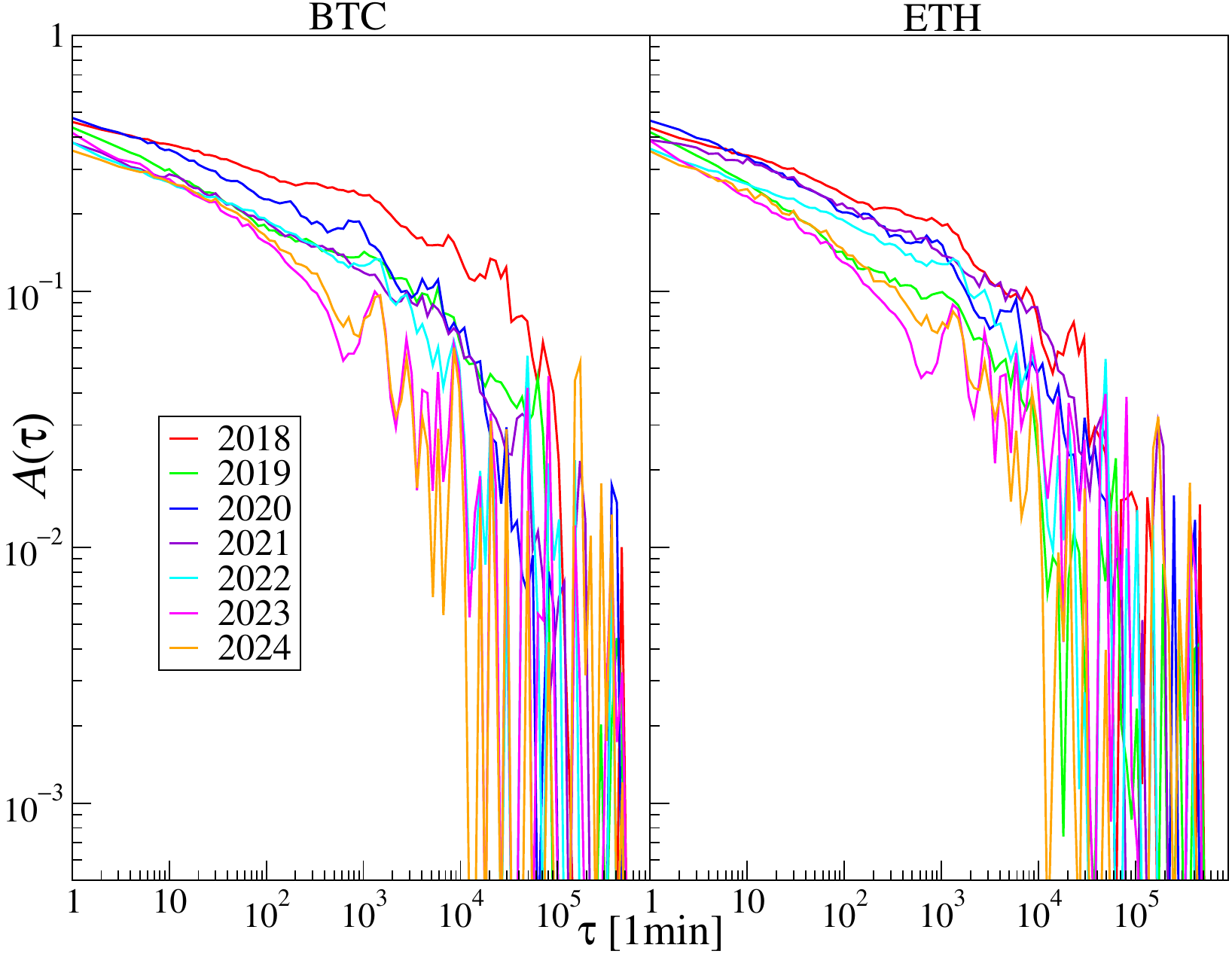}
\caption{The Pearson autocorrelation function calculated from the return moduli for BTC and ETH broken by year.}
\label{fig::ACF}
\end{figure}

Another basic characteristics of the return time series is the autocorrelation function (ACF) defined as:
\begin{equation}
A(\tau) = { 1/T \sum_{i=1}^{T-1} \left[R(t_i) - \langle R(t_i) \rangle_{t_i} \right] \left[ R(t_i+\tau) - \langle R(t_i) \rangle_{t_i} \right] \over \sigma^2_x},
\end{equation}
where $\sigma_x$ is the estimated standard deviation of the considered time series, $\langle \cdot \rangle$ represents the estimated mean, and $\tau$ is the time lag expressed in minutes. The ACF calculated from the moduli of log-returns for the same two series and likewise broken down by years are shown in log-log scale in Fig.~\ref{fig::ACF}. Clearly, they reveal long-range power-law correlations up to $10^3 - 10^4$ minutes, after which the power-law decay ends and the autocorrelation function falls to noise level. The fitting exponents of the autocorrelation functions for each year are presented in Tab/~\ref{data-spec}. Interestingly, the range of such correlations is seen to be larger in 2018 as compared to the more recent years~\cite{KwapienJ-2022a}. This may be related to the fact that, over the years, transaction frequency has been increasing on the BTC and ETH markets. This frequency together with a larger information inflow on the market have been a principal factor behind a well-documented observation that some market characteristics like the autocorrelation decay time or the log-return pdf tail index behave as if time flow accelerated when going from past to present~\cite{DrozdzS-2003a,DrozdzS-2007a}. This time flow may be identified as an internal market clock which operates at a non-uniform pace; the distinction between it and universal time is a well known concept in financial econometrics~\cite{clark1973}. (When time becomes to flow faster, everything that used to take place at a given characteristic time scale with a given intensity before, can now be observed with the same intensity at a shorter time scale -- for example, the autocorrelation function reaching noise level.)


\begin{table}[]
\caption{Basic statistics of the BTC and ETH exchange rate time series considered in this study: the average inter-transaction time $\delta t$, the power-law exponent $\gamma$ and stretched exponential function parameter $\beta$ least-square-fitted to the empirical return distributions presented in Fig.~\ref{fig::returns.pdf} and to empirical autocorrelation functions presented in Fig.~\ref{fig::ACF}.}
\begin{tabular}{|l|ll|ll|ll|}
\hline
              \textbf{year}& \multicolumn{2}{l|}{\textbf{$\langle \delta t \rangle$ [s]}}            & \multicolumn{2}{l|}{\textbf{$P(X>|x|)$ exponents}} & \multicolumn{2}{l|}{\textbf{ACF $\gamma$ exponents}}                \\ \hline
              & \multicolumn{1}{l|}{\textbf{BTC}} & \textbf{ETH} & \multicolumn{1}{l|}{\textbf{BTC}}   & \textbf{ETH}   & \multicolumn{1}{l|}{\textbf{BTC}} & \textbf{ETH} \\ \hline
\textbf{2018} & \multicolumn{1}{l|}{0.361}        & 0.604        & \multicolumn{1}{l|}{$\beta=0.48$}           & $\beta=0.48$          & \multicolumn{1}{l|}{-0.10}        & -0.13        \\ \hline
\textbf{2019} & \multicolumn{1}{l|}{0.241}        & 0.562        & \multicolumn{1}{l|}{$\gamma=-2.44$, $\beta=0.36$ }          & $\gamma=-2.58$, $\beta=0.39$          & \multicolumn{1}{l|}{-0.18}        & -0.23        \\ \hline
\textbf{2020} & \multicolumn{1}{l|}{0.101}        & 0.257        & \multicolumn{1}{l|}{$\gamma=-2.23$}          & $\gamma=-2.18$          & \multicolumn{1}{l|}{-0.16}        & -0.18        \\ \hline
\textbf{2021} & \multicolumn{1}{l|}{0.047}        & 0.065        & \multicolumn{1}{l|}{$\gamma=-2.58$}          & $\gamma=-2.66$          & \multicolumn{1}{l|}{-0.16}        & -0.14        \\ \hline
\textbf{2022} & \multicolumn{1}{l|}{0.026}        & 0.096        & \multicolumn{1}{l|}{$\gamma=-2.83$}          & $\gamma=-2.76$, $\beta=0.39$           & \multicolumn{1}{l|}{-0.16}        & -0.15        \\ \hline
\textbf{2023} & \multicolumn{1}{l|}{0.034}        & 0.145        & \multicolumn{1}{l|}{$\gamma=-2.44$}          & $\gamma=-2.37$          & \multicolumn{1}{l|}{-0.23}        & -0.25        \\ \hline
\textbf{2024} & \multicolumn{1}{l|}{0.045}        & 0.069        & \multicolumn{1}{l|}{$\gamma=-3.12$}          & $\gamma=-3$          & \multicolumn{1}{l|}{-0.19}        & -0.21        \\ \hline
\end{tabular}
\label{data-spec}
\end{table}

\subsection{Multifractal formalism}
\label{sect::multifractal.formalism}

Among the various methods available for fractal analysis of time series, MFDFA~\cite{KantelhardtJ-2002a} has been recognized as one of the most robust and reliable techniques~\cite{OswiecimkaP-2006a}. This method is specifically designed to handle nonstationary data by systematically removing trends across multiple time scales and analyzing the statistical characteristics of the resulting fluctuations. An extension of this approach, known as the multifractal cross-correlation analysis (MFCCA~\cite{OswiecimkaP-2014a}), enables the detection of multiscale cross-correlations between two concurrent nonstationary time series~\cite{PodobnikB-2008a,ZhouWX-2008a,HorvaticD-2011a}. In the following, we provide a brief overview of the MFCCA methodology.

Consider two nonstationary time series, ${\rm X} = \{X_i\}_{i=1}^T$ and $Y = \{Y_i\}_{i=1}^T$, uniformly sampled at intervals of $\Delta t$. To begin the analysis, each series is divided into $M_s = 2 \lfloor T/s \rfloor$ non-overlapping segments of length $s$, where the division is performed from both the beginning ($i=1$) and the end ($i=T$) of the series. Here, $\lfloor \cdot \rfloor$ denotes the floor function. Within each segment, the time series is integrated, and a polynomial trend $P_{\nu}^m(j)$ of degree $m$ is subsequently removed from the integrated signal:
\begin{equation}
x_j(s,\nu) = \sum_{k=1}^j X_{j(\nu-1)+k} - P_{\nu}^m(j), \quad j=1,...,s, \quad \nu=1,...,M_s.
\label{eq::detrended profile}
\end{equation}
Usually, a polynomial of degree $m=2$ constitutes a reasonable choice~\cite{OswiecimkaP-2013a}. Thus, in each segment, a detrended covariance:
\begin{equation}
f_{\rm XY}^2(s,\nu) = {1 \over s} \sum_{j=1}^s \left[ x_j(s,\nu) - \langle x_j(s,\nu) \rangle_j \right] \left[ y_j(s,\nu) - \langle y_j(s,\nu) \rangle_j \right],
\label{eq::detrended.covariance}
\end{equation}
is calculated, where $\langle \cdot \rangle_j$ denotes the averaging over $j$. The covariances in all segments and then the signed moments of order $r$ are calculated. These are called the bivariate fluctuation functions of $s$:
\begin{equation}
F_r^{\rm XY}(s) = \big\{ {1 \over M_s} \sum_{\nu=1}^{M_s} {\rm sign} [f_{\rm XY}^2(s,\nu)] | f_{\rm XY}^2(s,\nu)|^{r/2} \big\}^{1/r}.
\label{eq::fluctuation.function.xy} 
\end{equation}
Because covariances can take negative values, using their absolute value ensures that $F_r^{\rm XY}$ remains real-valued, while the inclusion of the sign function preserves the consistency of the analysis~\cite{OswiecimkaP-2014a}. The nature of the functional dependence of $F_r^{\rm XY}$ on the scale $s$ enables one to differentiate between fractal time series and those lacking such properties. Of particular interest are time series for which the fluctuation functions display power-law scaling over a sufficient range of moments $r$ and scales $s$:
\begin{equation}
F_r^{\rm XY}(s) \sim s^{\lambda(r)},
\end{equation}
Here, $\lambda(r)$ serves as the bivariate generalized Hurst exponent. In the case of monofractal cross-correlations, $\lambda(r)$ remains constant for all values of $r$, i.e., $\lambda(r) = \mathrm{const}$. Conversely, multifractal cross-correlations are characterized by a monotonically decreasing $\lambda(r)$ as a function of $r$.

A special case of $F_r^{\rm XY}$ arises when ${\rm X} = {\rm Y}$, in which the detrended cross-correlations reduce to detrended autocorrelations. In this situation, both the sign function and the absolute value in Eq.~(\ref{eq::fluctuation.function.xy}) can be omitted, as the detrended variance $f_{\rm X}^2$ is always non-negative. Consequently, the MFCCA reduces to the standard MFDFA, yielding univariate fluctuation functions $F_r^{\rm XX}(s)$ and $F_r^{\rm YY}(s)$. Here, the detrended cross-correlation function represents the mean covariance, while $F_r^{\rm XX}$ and $F_r^{\rm YY}$ correspond to the mean variances. If these univariate fluctuation functions follow a power-law scaling with respect to $s$, such that
\begin{equation}
F_r^{\rm XX}(s) \sim s^{h_{\rm X}(r)} \\\\\\\\\ \ \ \ \ \ \ \
F_r^{\rm YY}(s) \sim s^{h_{\rm Y}(r)},
\end{equation}
the exponents $h_{\rm X}(r)$ and $h_{\rm Y}(r)$ are identified as the generalized Hurst exponents. For $r = 2$, they reduce to the classical Hurst exponent $H$.

A standard approach to characterize the multifractality of data is through the singularity spectrum $f(\alpha)$. It is obtained from $h(r)$ via the Legendre transform:
\begin{eqnarray}
\nonumber
\alpha &=& h(r) + r h'(r),\\
f(\alpha) &=& r \left[\alpha - h(r)\right] + 1,
\label{eq::singularity.spectrum}
\end{eqnarray}
where $\alpha$ quantifies the intensity of a local singularity and corresponds to the H\"older exponent~\cite{HalseyTC-1986a}. Geometrically, $f(\alpha)$ represents the fractal dimension of the subset of the data whose H\"older exponent equals $\alpha$. For a monofractal time series, $(\alpha, f(\alpha))$ collapses to a single point, whereas for a multifractal, it forms a concave curve with downward-pointing shoulders. The broader the spectrum, the stronger the multifractality, making it a useful measure of time series complexity. Its quantitative extent is often expressed by the spectrum width:
\begin{equation}
\Delta\alpha := \alpha_{\rm max} - \alpha_{\rm min} = \alpha(r_{\rm min}) - \alpha(r_{\rm max}).
\label{eq::falpha.width}
\end{equation}
Alternatively, these properties can be described by the scaling function $\tau(r)$, defined as
\begin{equation}
\tau(r) = rh(r) - 1.
\end{equation}
In the monofractal case, $\tau(r)$ varies linearly with $r$ (since $h(r)$ is constant), whereas for multifractal series, it becomes nonlinear.

While for synthetic series $f(\alpha)$ is typically symmetric, in realistic cases it may also appear distorted and asymmetric, indicating that data points of different magnitudes possess distinct hierarchical organization~\mbox{\cite{OhashiK-2000a,CaoG-2013a,DrozdzS-2015a,GomezGomezJ-2021a}}.
Thus, another characteristic of the multifractal spectrum is its asymmetry, which reveals additional aspects of the data temporal organization. The related asymmetry parameter is defined~\cite{DrozdzS-2015a} as
\begin{equation}
A_\alpha=(\Delta\alpha_\textrm{L}-\Delta\alpha_\textrm{R})/(\Delta\alpha_\textrm{L}+\Delta\alpha_\textrm{R})\textrm{  ,}
\label{eq::falpha.asymmetry}
\end{equation}
where $\Delta\alpha_\textrm{L}=\alpha_0-\alpha_\textrm{min}$ and $\Delta\alpha_\textrm{R}=\alpha_\textrm{max}-\alpha_0$. $\alpha_0$ denotes $\alpha$ for which $f(\alpha)$ reaches maximum. 
Left-sided asymmetry indicates that multifractality is dominated more by large events and the small ones are contracted more towards noise. The opposite applies to right-sided asymmetry where multifractality is dominated by small events; large ones lose hierarchical organization.

\subsection{Detrended cross-correlation coefficient}
\label{sect::rhor}

Having calculated all fluctuation functions, one can then also introduce the $q$-dependent detrended cross-correlation coefficient $\rho_q(s)$~\cite{kwapien2015} defined as
\begin{equation}
\rho_r(s) = {F_r^{\rm XY}(s) \over \sqrt{ F_r^{\rm XX}(s) F_r^{\rm YY}(s) }}.
\label{eq::rho.r}
\end{equation}
This quantity can be considered as the counterpart of the Pearson cross-correlation coefficient for non-stationary signals. Both coefficients assume values in the range [-1,1] with $\rho_r(s)=1$ for perfectly correlated time series, $\rho_r(s)=0$ for independent time series, and $\rho_r(s)=-1$ for perfectly anticorrelated time series. It should be noted that, in order to calculate $\rho_r(s)$, the time series do not have to be fractal~\cite{kwapien2015}.

\subsection{Decomposing sources of multifractality}
\label{sect::sources}

Obtaining reliable numerical outcomes using the MFDFA algorithm described above is inherently challenging. The risk of overestimation and misinterpretation is high, particularly when working with relatively short time series. Standard techniques for generating free surrogate data — such as shuffling original series with only a few thousand data points — may misleadingly exhibit multifractal scaling. In such cases, the apparent multifractality arises not from genuine structural features, but from residual correlations~\cite{KwapienJ-2023a}. This issue becomes especially prominent when the time series displays heavy-tailed fluctuation distributions as in the present case of digital market instruments. In these scenarios, only sufficiently long datasets can reveal the true absence of multifractality. In practice, short time series with heavy-tailed fluctuations often produce spurious indications of multifractal behavior~\cite{DrozdzS-2009a}. A recent approach~\cite{KluszczynskiR-2025a} offers a more systematic validation of multifractality by attenuating heavy-tailed fluctuations while preserving correlation structures. This method involves a ranking-based probability density transformation that reprojects the data, retaining their temporal order while progressively reshaping the fluctuation distribution toward a Gaussian form~\cite{KluszczynskiR-2025a}.

$q$-Gaussian distributions provide a highly practical analytical framework for modeling this class of probability distributions. They serve as a natural extension of the standard Gaussian distribution, analogous to how the Tsallis entropy $S_q$ extends the classical Boltzmann—Gibbs entropy $S$~\cite{UmarovS-2008a,TsallisC-1995a}. The $q$-Gaussian distribution $\mathcal{G}q$ is characterized by two parameters: a shape parameter $q \in (-\infty, 3)$ and a scale (width) parameter $\beta > 0$. Its probability density function (PDF) takes the form~\cite{UmarovS-2008a}:
\begin{equation}
p_q(x) = \frac{\sqrt{\beta}}{C_q} e_q(-\beta x^2),
\label{eq::qgaussian.pdf}
\end{equation}
where $e_q(x)$ denotes the $q$-exponential function, defined as
\begin{equation}
e_q(x) = \begin{cases}
(1+(1-q)x)^{1/(1-q)} & \text{if } q\neq 1 \text{ and } 1+(1-q)x>0\\
0 & \text{if } q\neq 1 \text{ and } 1+(1-q)x\leq 0\\
e^x & \text{if } q=1,
\end{cases}
\end{equation}
and $C_q = \int{-\infty}^{\infty} e_q(x^2),dx$ is a normalization constant.

The persistence of well-defined multifractality characteristics in the limit $q=1$, which corresponds to the Gaussian distribution, is a strong indicator that the observed multifractality stems from genuine correlations. It is expected and important to note that heavy-tailed distributions can enhance multifractality, but only when temporal correlations are present. In the absence of such correlations, the system typically exhibits monofractal behavior, or in more extreme cases, bifractality, particularly when the fluctuation dynamics fall within the L\'evy-stable regime~\cite{KwapienJ-2023a}.

\section{Results}
\label{sect::results}

\subsection{Multifractal characteristics of BTC and ETH in the years 2018-2024}
\label{sect::BTC.ETH.univariate}

\begin{figure}[ht!]
\centering
\includegraphics[width=0.99\textwidth]{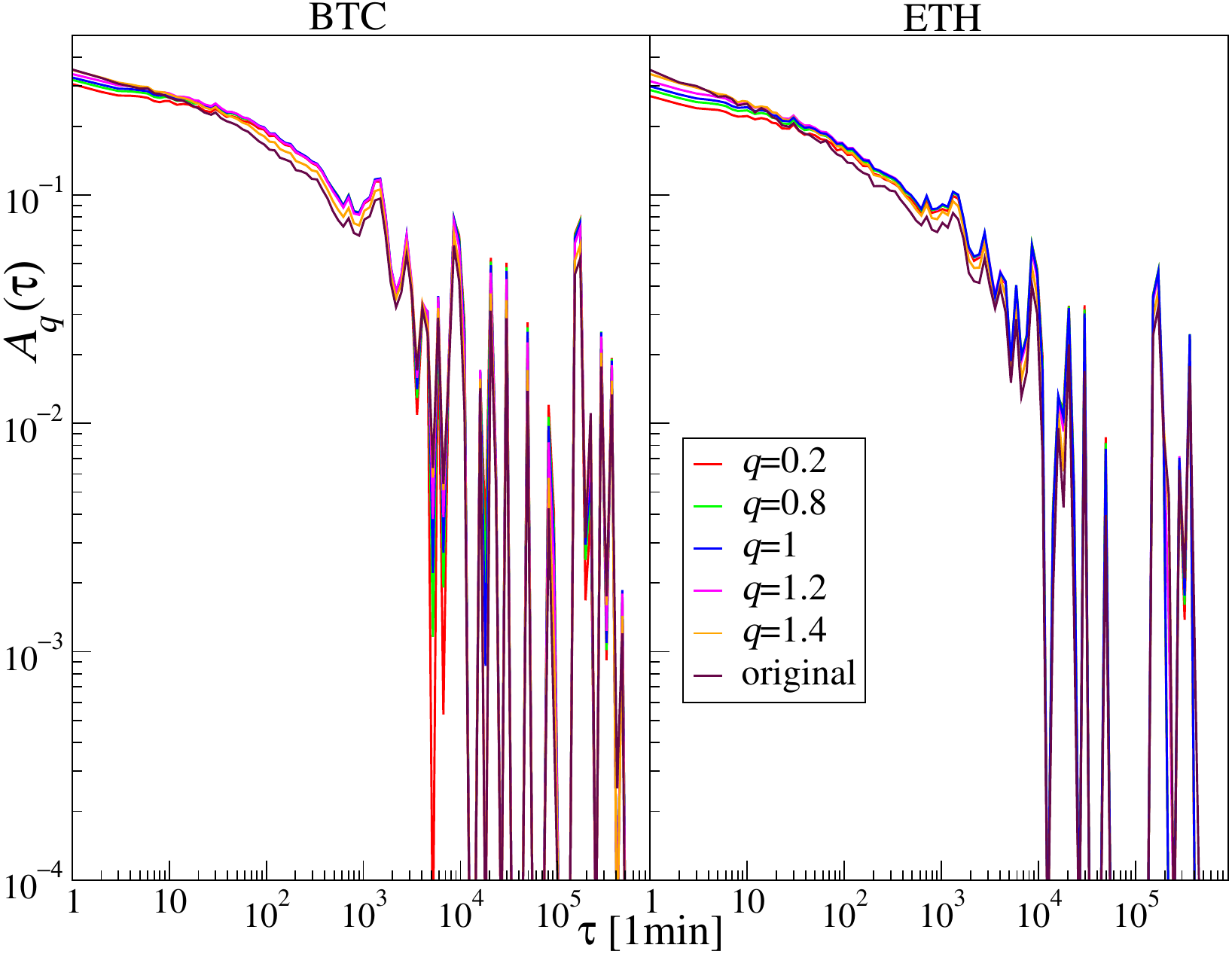}
\caption{The Pearson autocorrelation function calculated from the return moduli for BTC and ETH in 2024 with their original PDFs replaced by the $q$-Gaussian distributions with different values of $q$.}
\label{fig::ACF24}
\end{figure}

The validity of the assumption about the preservation of temporal correlations when converting fluctuation distributions to distributions with varying tail thickness, as modeled here with the $q$-Gaussians, is illustrated in Fig.~\ref{fig::ACF24} for the year 2024 as an example from Fig.~\ref{fig::ACF}. Indeed, starting from empirical data (best fitted with $q \approx 1.5$) and then projecting then onto $q$-Gaussians for $q$ staring from $q=1.4$ towards a Gaussian $(q=1)$ and further down to $q=0.2$, which corresponds to an almost uniform distribution, it is clearly seen that the form of $A(\tau)$ remains practically unchanged. This even includes a sharp drop to zero around the $10^3-10^4$ timescale, meaning that for larger time distances, fluctuations become uncorrelated. Interestingly, this process begins slightly earlier for higher capitalization BTC (slightly above $10^3$) than for ETH (closer to $10^4$). From a more general perspective these results provide a further strong argument for the validity of the methodology proposed by Kluszczyński \textit{et al.}~\cite{KluszczynskiR-2025a} that temporal correlations measured by the Pearson coefficient and distributions of fluctuations can be disentangled.


\begin{figure}[ht!]
\centering
\includegraphics[width=0.99\textwidth]{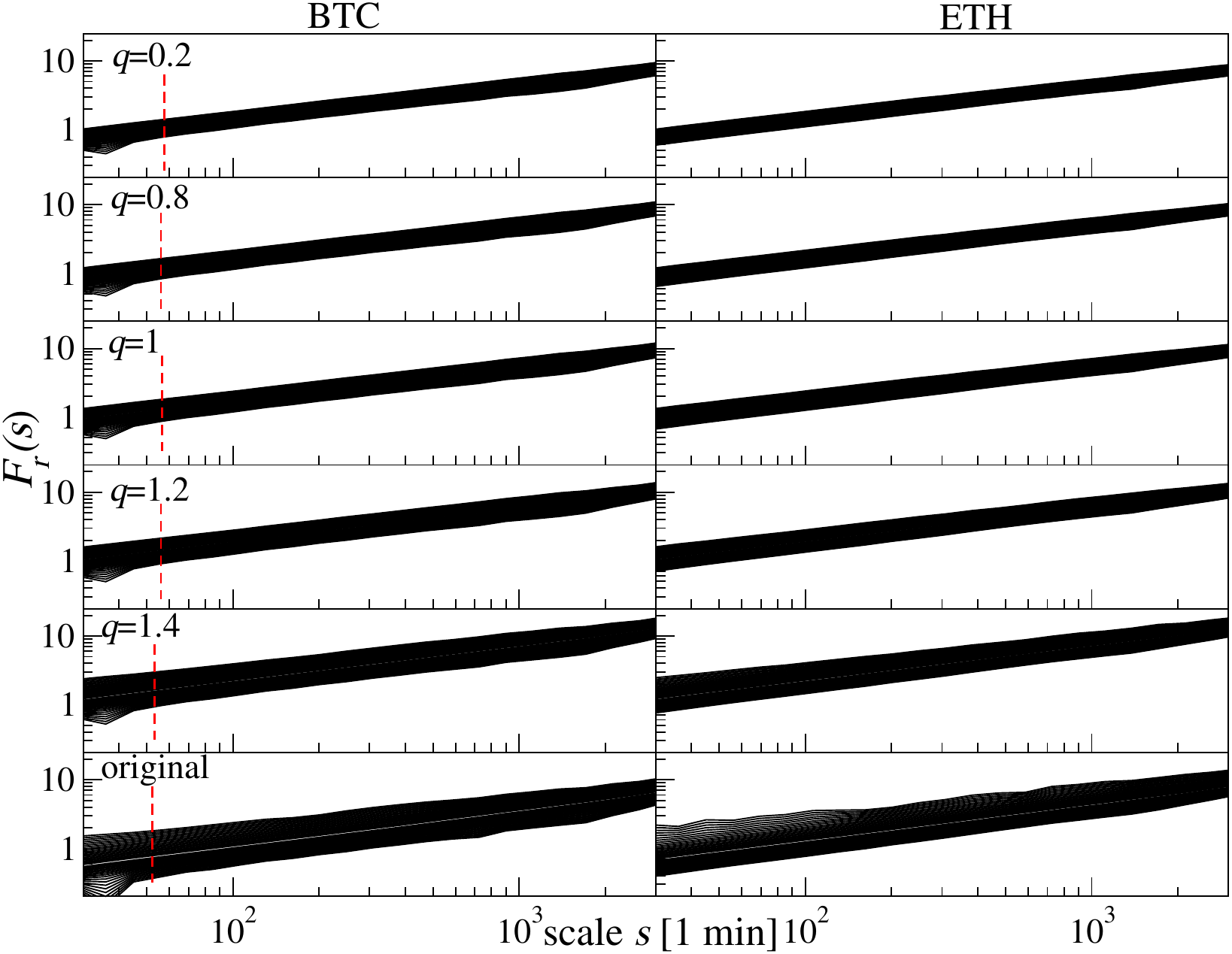}
\caption{Fluctuation functions $F_r(s)$ for BTC and ETH in 2024 with their original PDFs replaced by the $q$-Gaussian distributions with different values of $q$. Start of the scaling range, in which a power-law form of $F_r(s)$ is observed for a range of values of $q$, is denoted by a vertical red dashed line.}
\label{fig::Fr24}
\end{figure}

At the same time, the corresponding univariate fluctuation functions calculated for $-4 \leq r \leq 4$ for the same 2024 year and shown in Fig.~\ref{fig::Fr24} display scaling for both BTC and ETH. This scaling is $r$-dependent, though not very strongly, for $q = 1$ and smaller, to become clearly multifractal already for $q = 1.2$, and even more so for $q=1.4$, with the strongest $q$-dependence for the empirical data considered. It should be noted, however, that in the case of BTC, the uniform scaling of $F_r(s)$ breaks down slightly around $2\times 10^3$, while for ETH the uniform scaling persists over the entire interval $s$ shown. Such observations correspond well to the relative behavior of the autocorrelation function $A(\tau)$ discussed above, where the one for ETH remains positive for a longer period.


\begin{figure}[ht!]
\centering
\includegraphics[width=0.99\textwidth]{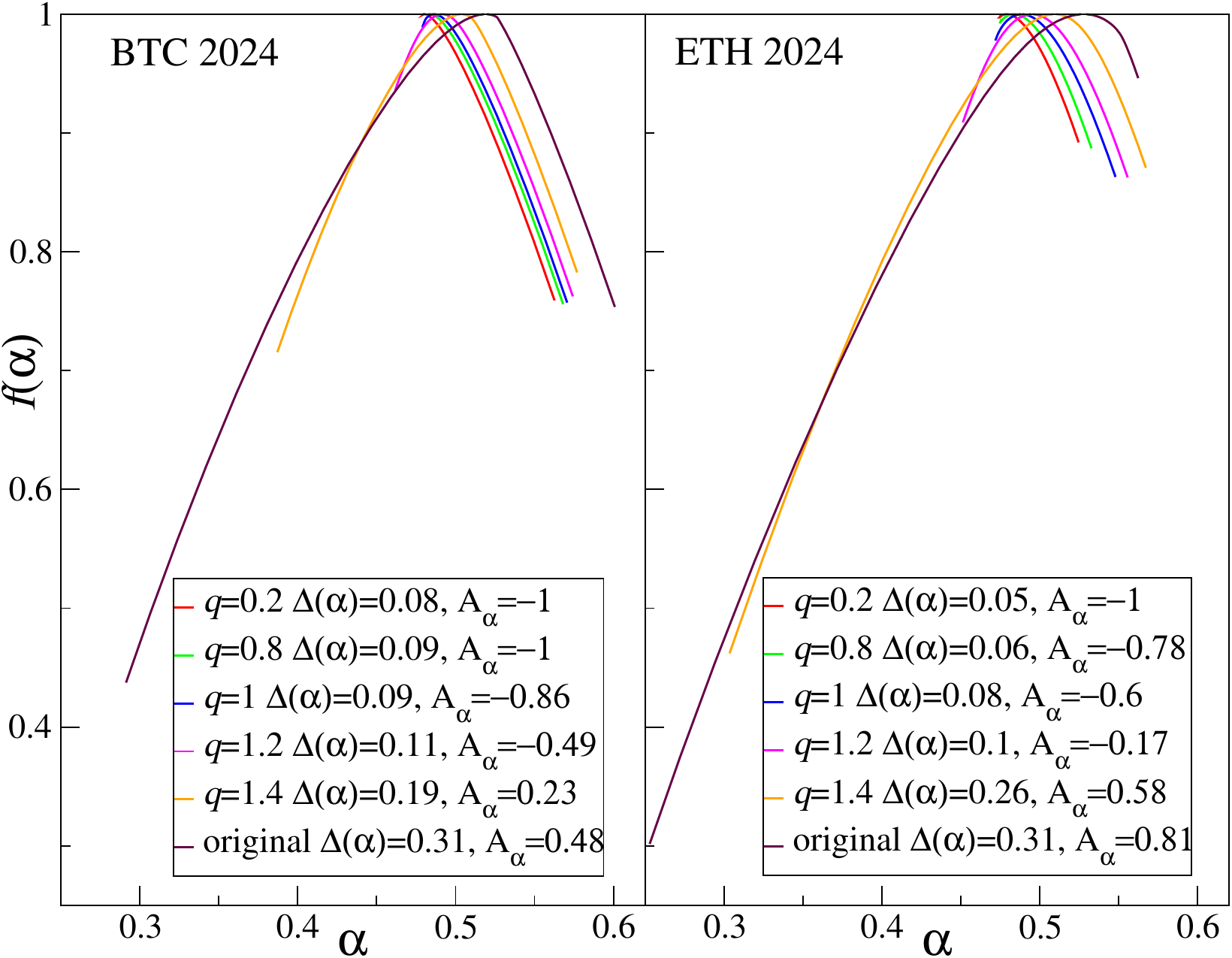}
\caption{Multifractal spectra for BTC and ETH in 2024 with their original PDFs replaced by the $q$-Gaussian distributions with different values of $q$.}
\label{fig::spectra24}
\end{figure}

The scaling characteristics are most clearly revealed in the singularity spectrum representation (also called multifractal spectrum) which, for the cases shown in Fig.~\ref{fig::Fr24} are presented in Fig.~\ref{fig::spectra24}. A distinct multifractal shape is visible already at the smallest analyzed values of $q$, as reflected by the widths of the $\Delta (\alpha)$ spectra. They change very slowly as $q$ increases from 0.2 to 1.0, after which they begin to grow increasingly faster, reaching values exceeding 0.3 for the empirical series. This increase is clearly caused by the increasing thickness of the tails of the PDFs. However, it should be emphasized again that this result is possible because there are temporal correlations in the analyzed series, which are visible in the autocorrelation functions above. Destroying these correlations, for example, by reshuffling the time series reduces these spectra to points, which indicates monofractality. These effects can be seen here because the considered time series are sufficiently long to ensure convergence~\cite{DrozdzS-2009a,ZhouWX-2012a,KwapienJ-2023a}. Thus, the thickness of the PDF tails has a large constructive importance for the width of multifractal spectra, but without time correlations, multifractality has no support to exist. Another related significant effect, which draws attention, is the asymmetry of the multifractal spectra as defined by Eq.~(\ref{eq::falpha.asymmetry}). The corresponding numbers are listed in Fig.~\ref{fig::spectra24}. The spectra for the original empirical data have the most pronounced left-side character, which is rather typical for financial series and reflects the fat-tailed shape of the corresponding PDFs. In both cases, the asymmetry parameter $A_{\alpha}$ is relatively large, but it is even significantly larger for ETH than for BTC (0.48 vs. 0.81). This corresponds to the thicker tails of the log-return distributions in the ETH case and is consistent with the higher frequency of extreme events in 2024 visible in Fig.~\ref{fig::returns.pdf}. (It is worthwhile to recall here the bifractal property of the L\'evy flight process pdfs that results in a two-point $f(\alpha)$ spectrum~\cite{NakaoH-2000a}. This property is sometimes ``sensed'' by data with thinner pdf tails, which leads to smearing of the singularity spectrum left shoulder over a disproportionately large range of $\alpha$'s~\cite{DrozdzS-2009a}.) At the same time, it can be seen that, by replacing the original empirical PDFs with $q$-Gaussians with decreasing $q$, $A_{\alpha}$ quickly and systematically decreases and it reaches a clearly negative value for $q=1$, indicating a right-sided asymmetry. Finally, for $q=0.2$, $A_{\alpha}$ takes on a value of -1, indicating the presence of only the right wing in $f(\alpha)$. Such a systematic change in the shape of the multifractal spectrum from the left-sided to the right-sided form with decreasing $q$ can actually be expected, because small $q$ correspond to increasingly more homogeneous PDFs and, therefore, to a lower probability of small values of the H\"older exponents determining $\alpha$.


\begin{figure}[ht!]
\centering
\includegraphics[width=0.99\textwidth]{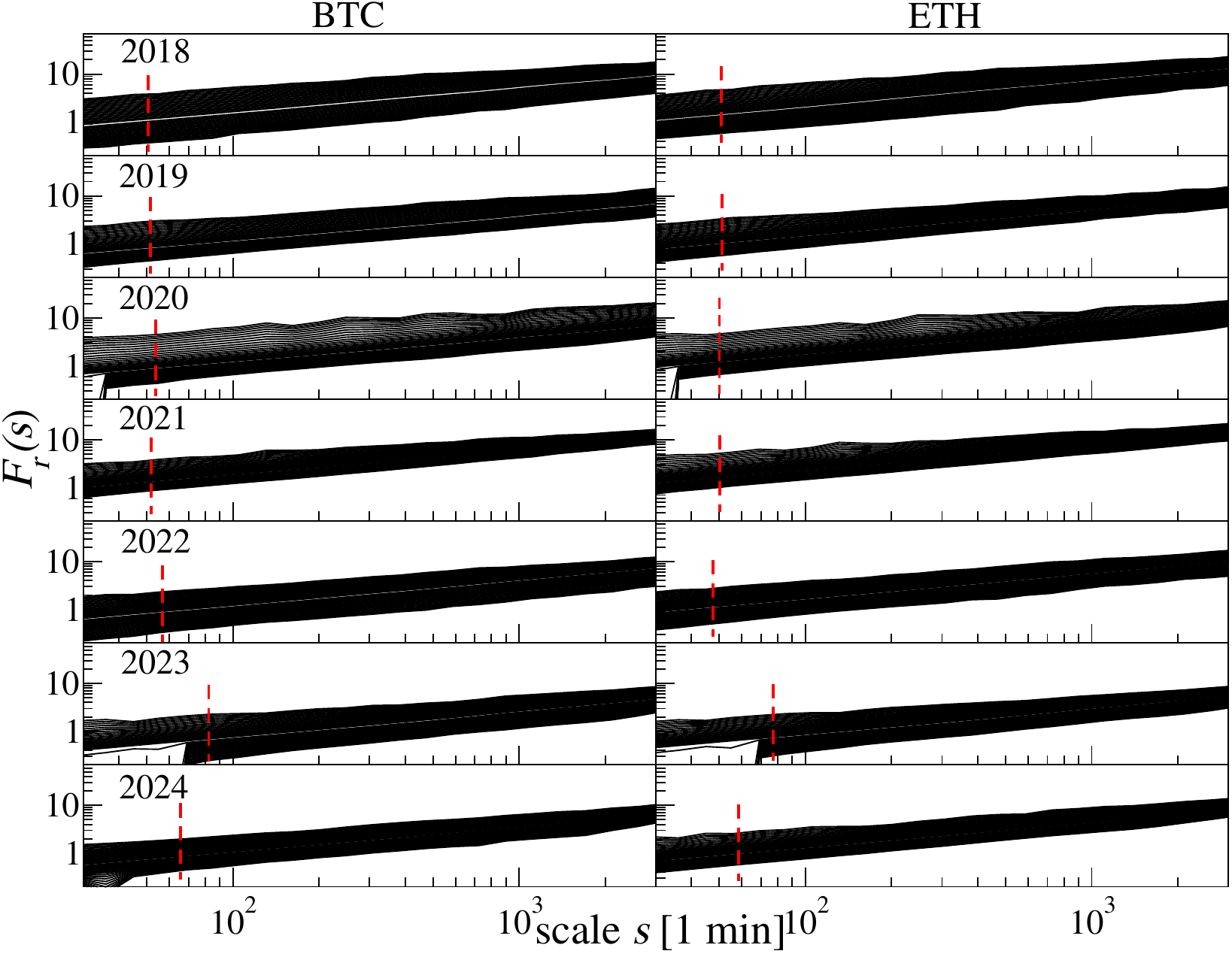}
\caption{Univariate fluctuation functions $F_r(s)$ for BTC and ETH across different~years. Start of the scaling range, in which a power-law form of $F_r(s)$ is observed for a range of values of $q$, is denoted by a vertical red dashed line.}
\label{fig::Fr}
\end{figure}

\begin{figure}[ht!]
\centering
\includegraphics[width=0.99\textwidth]{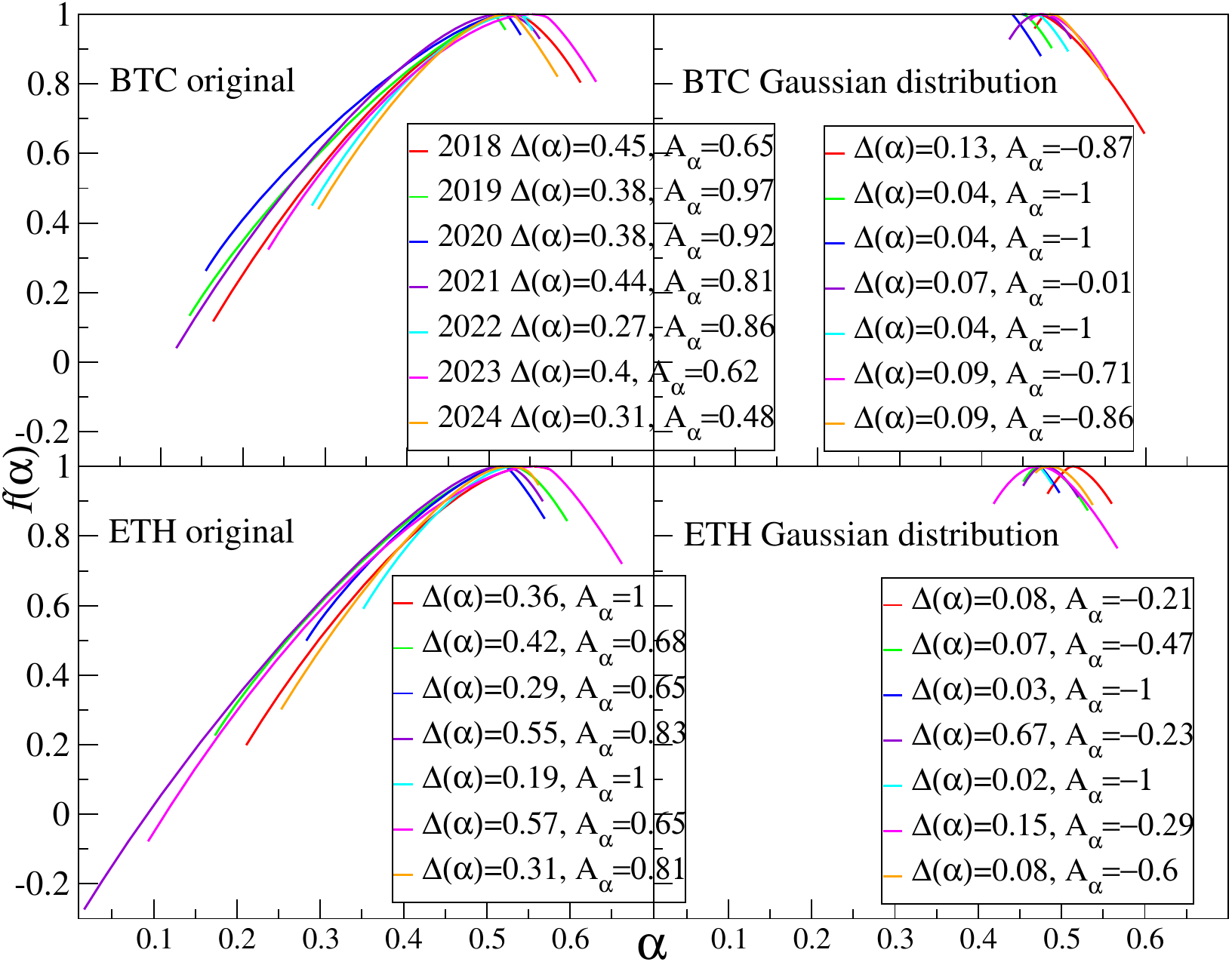}
\caption{Multifractal spectra for BTC and ETH across different years.}
\label{fig::spectra4}
\end{figure}

As it can be seen in Fig.~\ref{fig::returns.pdf}, the BTC and ETH PDFs calculated separately for the years 2018-2024 are dispersed quite clearly in relation to the inverse cubic power-law (black dashed line). The corresponding univariate fluctuation functions calculated for $-4 \leqslant r \leqslant 4$ for the same sequence of years are shown in Fig.~\ref{fig::Fr}. They display scaling for both assets, but with a slightly varying intensity of dependence on $r$ across different years. Of course, this also results in varying widths $\Delta\alpha$ of the multifractal spectra and their asymmetry coefficients, all displayed in Fig.~\ref{fig::spectra4}. $f(\alpha)$ corresponding to the Gaussianized PDFs ($q=1$), but with preserved temporal correlations, are also shown in the corresponding right panels. They are significantly narrower, indicating a significant role of the PDFs. However, it should be emphasized again that destroying temporal correlations by shuffling the time series reduces $\Delta\alpha$ to practically zero also for the original data — the ones with the broad distributions.


\subsection{Cross-correlations between BTC and ETH}
\label{sect::BTC.ETH.bivariate}

\begin{figure}[ht!]
\centering
\includegraphics[width=0.99\textwidth]{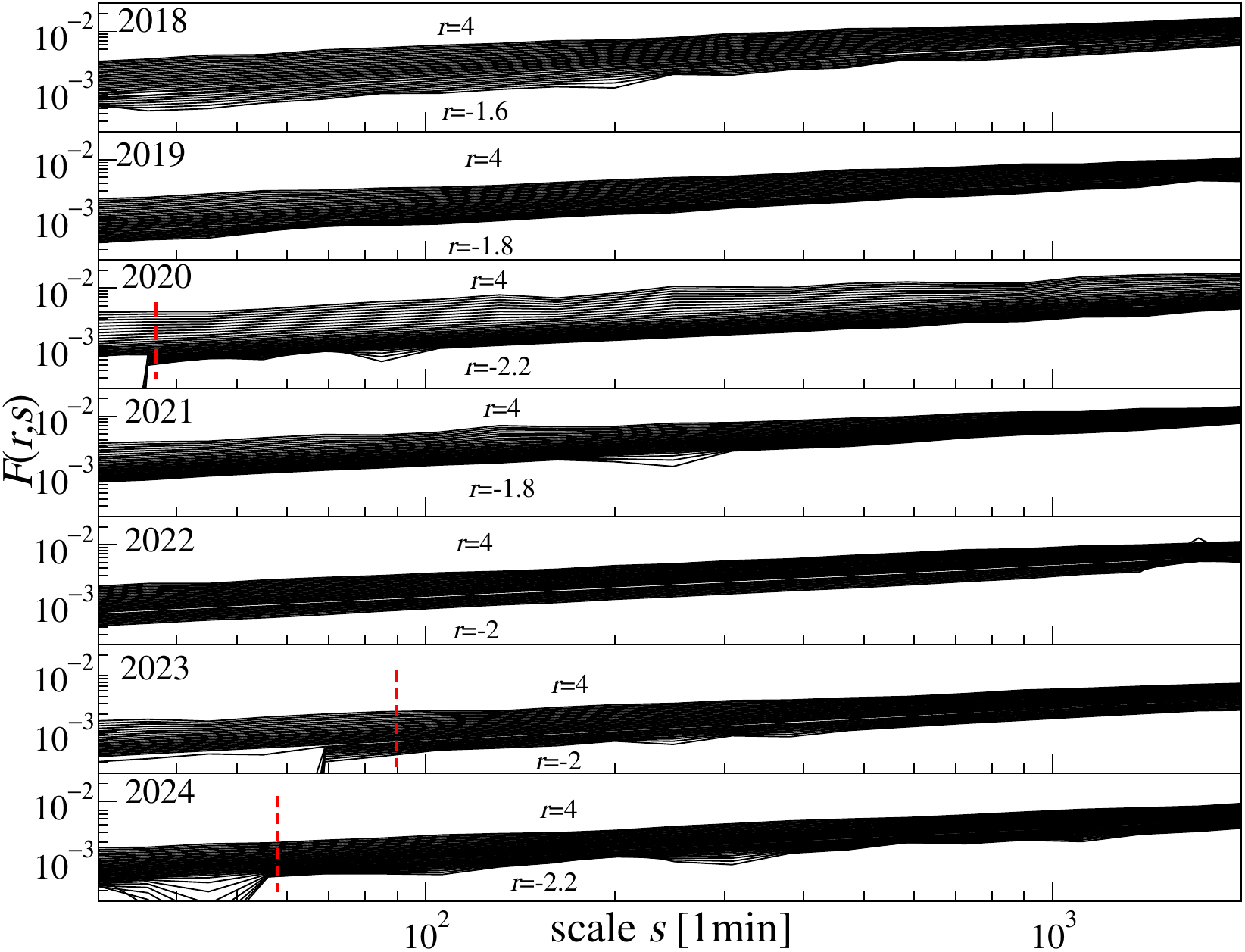}
\caption{Bivariate fluctuation functions $F_{\rm XY}(r,s)$ for BTC and ETH across different years. The~minimum values of $r$ vary among the years and are explicitly listed, while the maximum values are fixed at $r=4$. Start of the scaling range, in which a power-law form of $F_{\rm XY}(r,s)$ is observed for a range of values of $q$, is denoted by a vertical red dashed line (if applicable).}
\label{fig::FrXYorg}
\end{figure}

\begin{figure}[ht!]
\centering
\includegraphics[width=0.99\textwidth]{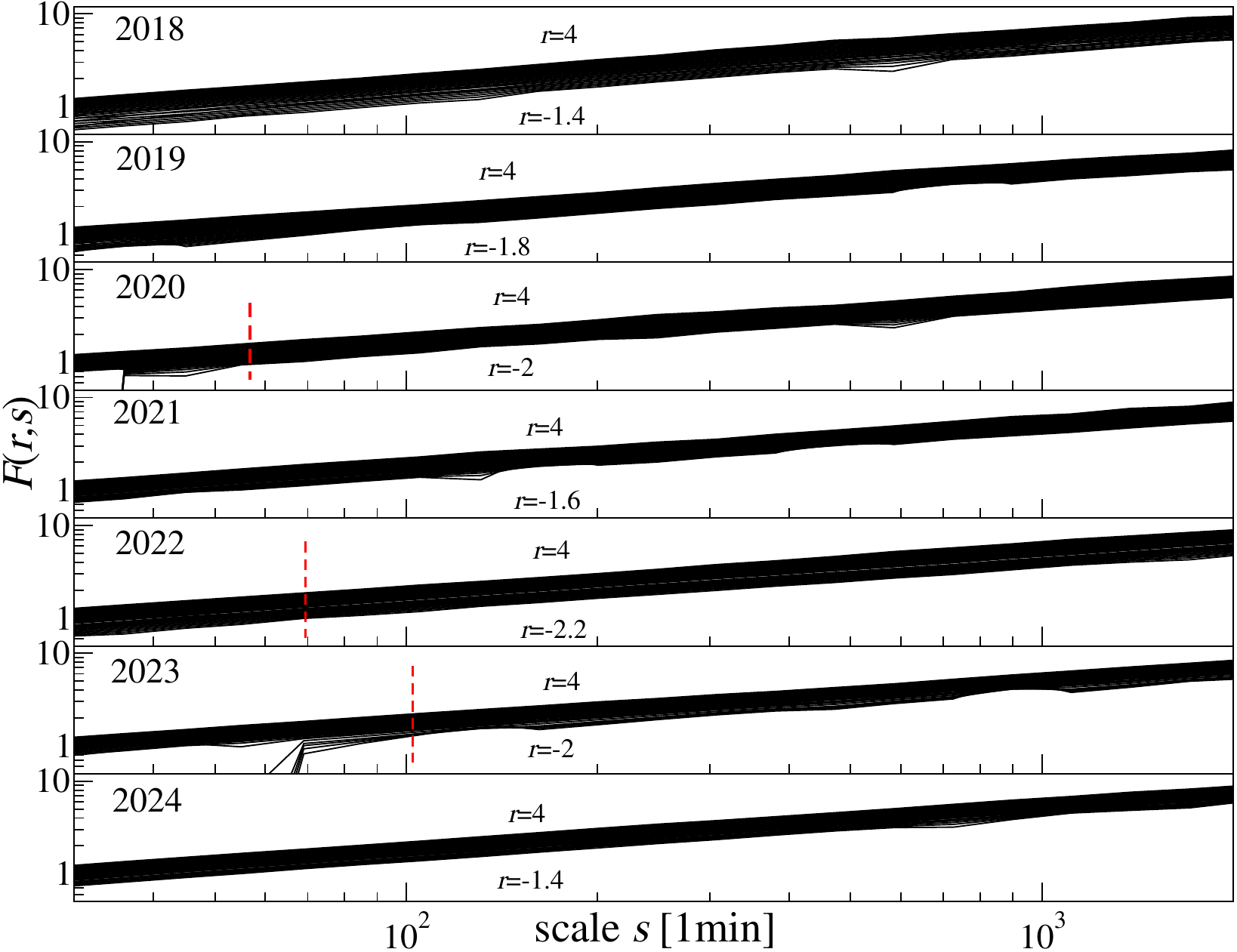}
\caption{Bivariate fluctuation functions $F_{\rm XY}(r,s)$ for BTC and ETH across different years after Gaussianizing the corresponding~PDFs. Start of the scaling range, in which a power-law form of $F_{\rm XY}(r,s)$ is observed for a range of values of $q$, is denoted by a vertical red dashed line (if applicable).}
\label{fig::FrXYG}
\end{figure}

The general MFCCA formalism outlined in Sect.~\ref{sect::multifractal.formalism}, via Eq.~(\ref{eq::detrended.covariance}) also allows us to quantitatively address the question of possible multifractal cross-correlations between two multifractal series, the ones representing the BTC and ETH returns in the present case. The corresponding cross-correlation functions are shown in Fig.~\ref{fig::FrXYorg}. They are displayed for those values of $r$ for which scaling of these functions is identifiable. As it can be seen, a relatively good scaling occurs for positive values of $r$, which amplify contribution from periods of large fluctuations. On the negative side of $r$, scaling typically terminates slightly below $r \approx -1.5$ for the data considered here. This means that, for the component corresponding to periods of small-amplitude noise, there are no fractal cross-correlations. An analogous result after Gaussianizing the same time series according to the procedure described in Sect.~\ref{sect::sources} gives the result shown in Fig.~\ref{fig::FrXYG}. The scaling is preserved over similar ranges of $r$, but it becomes definitely of the monofractal type, i.e. with a weak dependence on $r$.


\begin{figure}[ht!]
\centering
\includegraphics[width=0.99\textwidth]{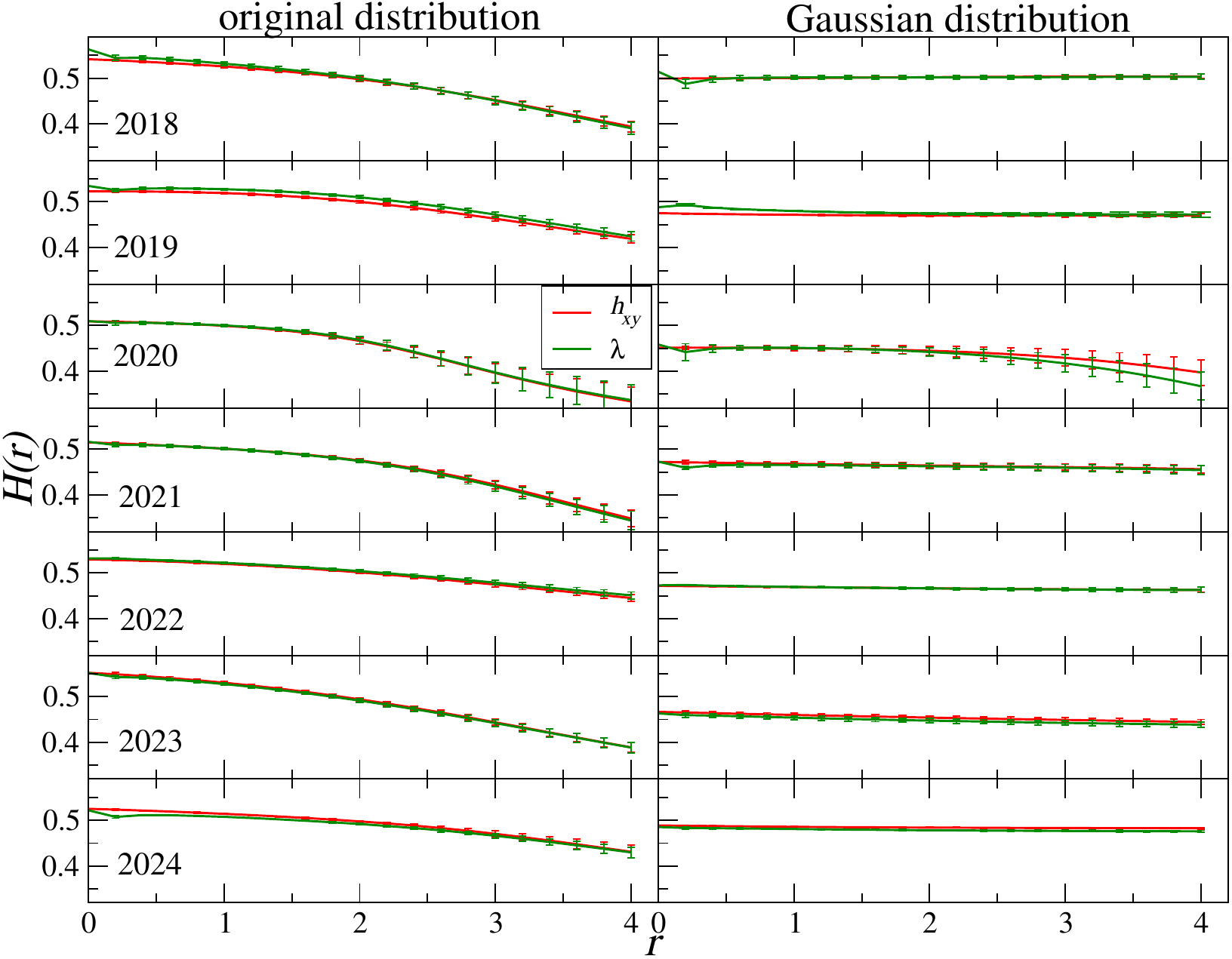}
\caption{The bivariate scaling exponent $\lambda(r)$ and the average generalized univariate Hurst exponent $h_{xy}(r)$ estimated from the fluctuations functions presented in Fig.~\ref{fig::FrXYorg} (left panels) and Fig.~\ref{fig::FrXYG} (right panels).}
\label{fig::lambda_sred}
\end{figure}

A fully quantitative estimation of the character of cross-correlations between time series is obtained in terms of the relation between $\lambda(r)$ and $h_xy(r) = (h_x(r) + h_y(r))/2$ introduced in Sect.~\ref{sect::multifractal.formalism}. The existence and similarity of these two quantities reflects the degree and quality of multifractal cross-correlations~\cite{OswiecimkaP-2014a}. For the BTC and ETH time series considered above, these two quantities are compared in Fig.~\ref{fig::lambda_sred} but only for $r \geqslant 0$, because for $r < 0$ the scaling is less pronounced or even disappears (see Fig~\ref{fig::FrXYorg}). From this perspective, a majority of the years 2018-2024 show convincing multifractal cross-correlations as both $\lambda(r)$ and $h_xy(r)$ are $r$-dependent and similar to each other. We remember, however, that this effect is visible for $r \geqslant 0$ corresponding to the periods of large fluctuations. The similarity between these two quantities is the weakest in the pandemic year 2020 due to larger estimation errors. This means that the time series, although each multifractal, were less cross-correlated. For completeness, Fig.~\ref{fig::FrXYorg} also shows the same quantities for the time series after Gaussianization of their PDFs ($q=1$). The $r$-dependence becomes much weaker here, although, admittedly, scaling still holds and both $\lambda(r)$ and $h_xy(r)$ can be determined. The lack of a clear dependence on $r$, however, indicates that the cross-correlation is monofractal here. It should be noted at the same time that the destruction of the autocorrelations by independent reshuffling of the BTC and ETH returns leads to a complete disappearance of any scaling associated with the cross-correlation.


\subsection{Detrended cross-correlation}
\label{sect::rhor.results}

\begin{figure}[ht!]
\centering
\includegraphics[width=0.99\textwidth]{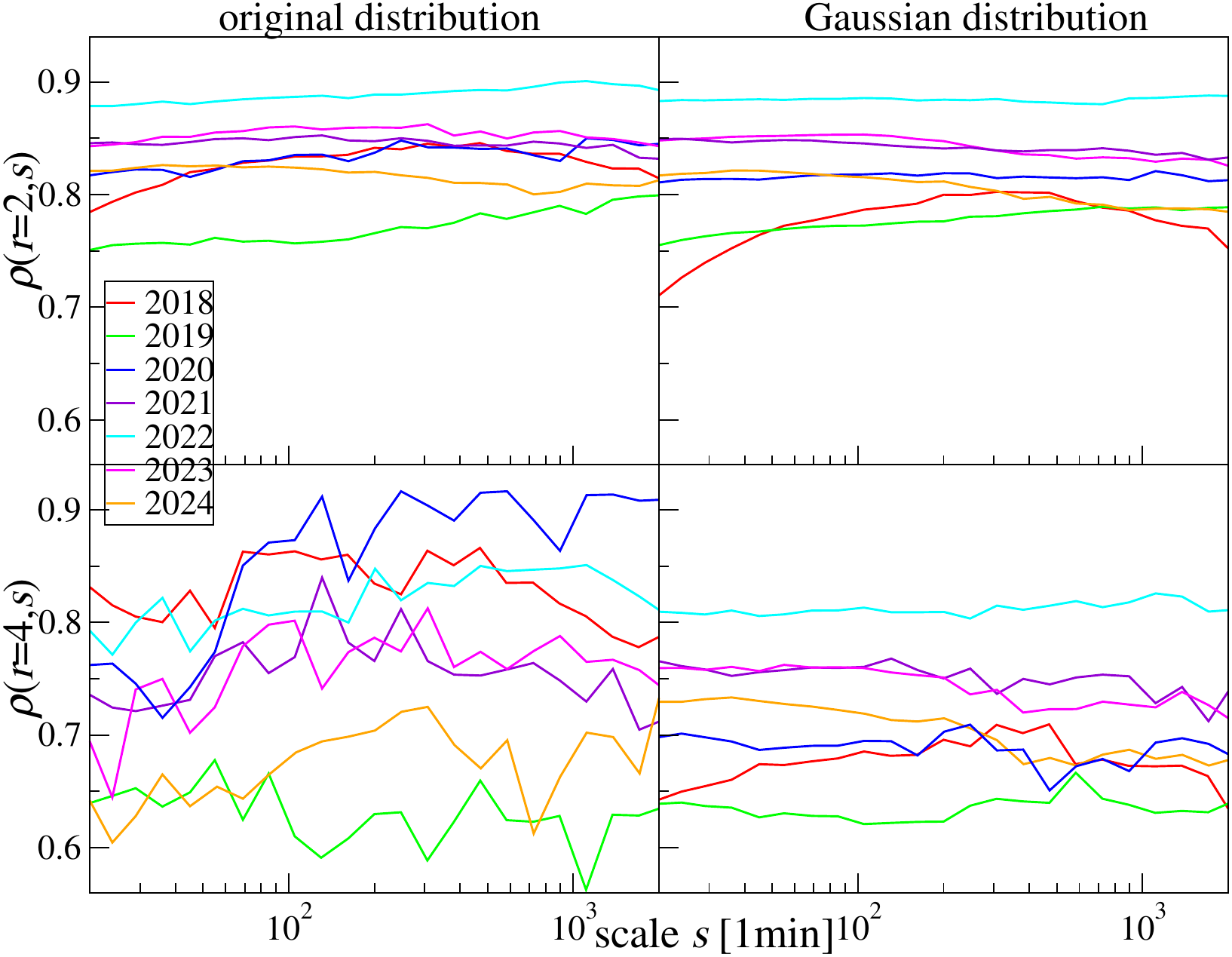}
\caption{Detrended cross-correlation coefficient for BTC and ETH across different years: the original time series (left panels) and after their Gaussianization (right panels).}
\label{fig::rhor.XY}
\end{figure}

A related but to some extent complementary description of the cross-correlation between the detrended time series can be obtained by using the coefficient $\rho_r(s)$ defined in Sect.~\ref{sect::rhor} by Eq.~(\ref{eq::rho.r}), which allows for resolving cross-correlations on a particular scale $s$ with respect to the size of the fluctuation amplitude. For the time series of BTC and ETH considered here, the values of $\rho_r(s)$ and their dependence on the scale $s$ for $r=2$ and $r=4$ are displayed in Fig.~\ref{fig::rhor.XY}. The former case, $r=2$, corresponds to the periods of typical medium-size fluctuations (which dominate the PDF) while the latter, $r=4$, retains the periods of more extreme fluctuations and, effectively, filters out both the periods of small fluctuations and the periods of average fluctuations. These coefficients are calculated for the original data (left side of Fig.~\ref{fig::rhor.XY}) as well as for their Gaussianized substitutes (right side). For the original data, the detrended cross-correlations are clear and quite similar in magnitude across the years considered. On average, they are slightly stronger for $r=2$ than for $r=4$, but in the latter case the dependence on year is much stronger. In particular, in 2020, the year of the Covid-19 pandemic, such cross-correlations at larger scales $s$ are even stronger for $r=4$ than for $r=2$, which seems understandable in light of the fact that price changes were more sudden then. Gaussianization significantly weakens this particular case, which confirms the role of large fluctuations in generating cross-correlations in 2020. In the other cases, Gaussianization generally weakens cross-correlations, although to a lesser extent.


\subsection{ETH from decentralized exchange}
\label{sect::DEX}

In contrast to traditional centralized exchanges (CEXs), such as Binance or Coinbase, where a single intermediary manages liquidity, order matching, and user access, decentralized exchanges (DEXs) like Uniswap rely on community-driven infrastructure and smart contracts to facilitate trading without intermediaries. While CEXs offer higher liquidity, ease of use, and established security procedures, DEXs provide greater anonymity and access to a wider set of tokens but often face challenges of lower liquidity, higher technical complexity, and exposure to contract-related risks. An analysis of Ethereum trading from June 2023 to June 2024~\cite{WatorekM-2024b} revealed that, despite these limitations, Uniswap already exhibited convincing signs of multifractality in its price and volume dynamics. However, unlike Binance, where multifractal patterns were more balanced and mature, the multifractal spectra on Uniswap were strongly left-skewed, indicating that multifractality originated primarily from periods of large fluctuations, with periods of small and average fluctuations behaving more like uncorrelated noise. Interestingly, the multifractality was more pronounced in transaction volumes than in returns, and cross-correlations between volatility and volume, though present, remained weaker than those observed on centralized platforms, underscoring the comparatively less mature yet evolving state of decentralized markets~\cite{WatorekM-2024b}.


\begin{figure}[ht!]
\centering
\includegraphics[width=0.49\textwidth]{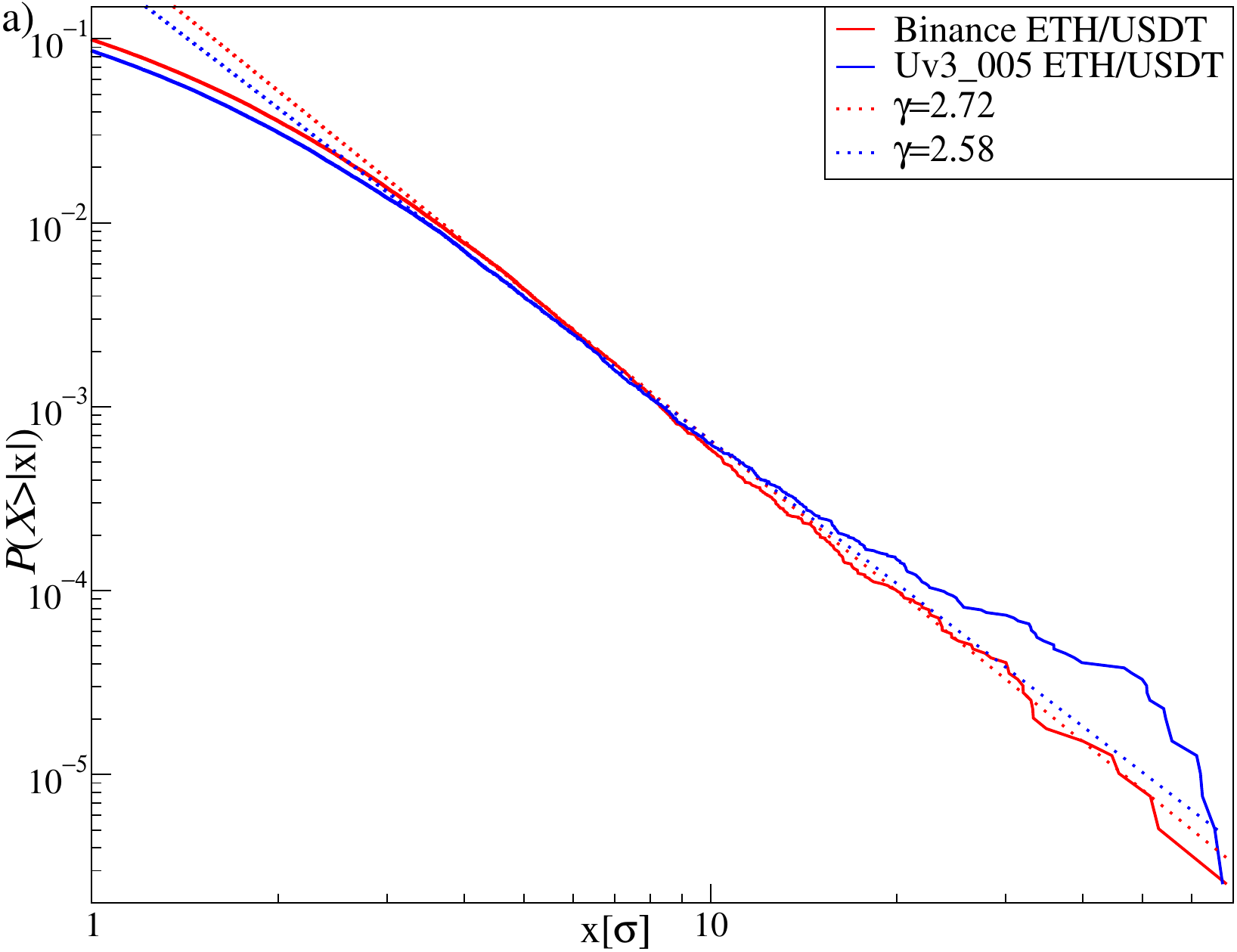}
\includegraphics[width=0.49\textwidth]{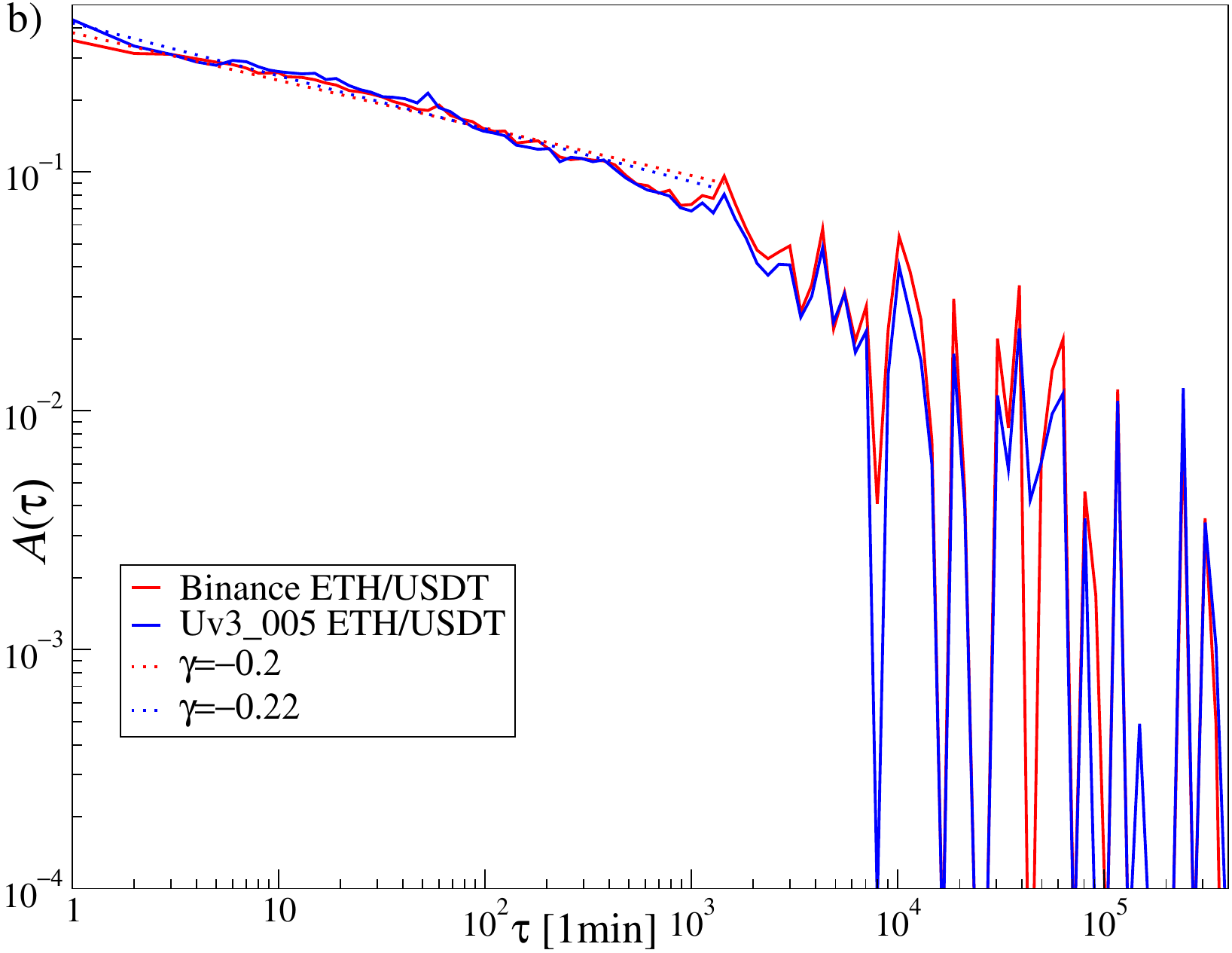}
\includegraphics[width=0.49\textwidth]{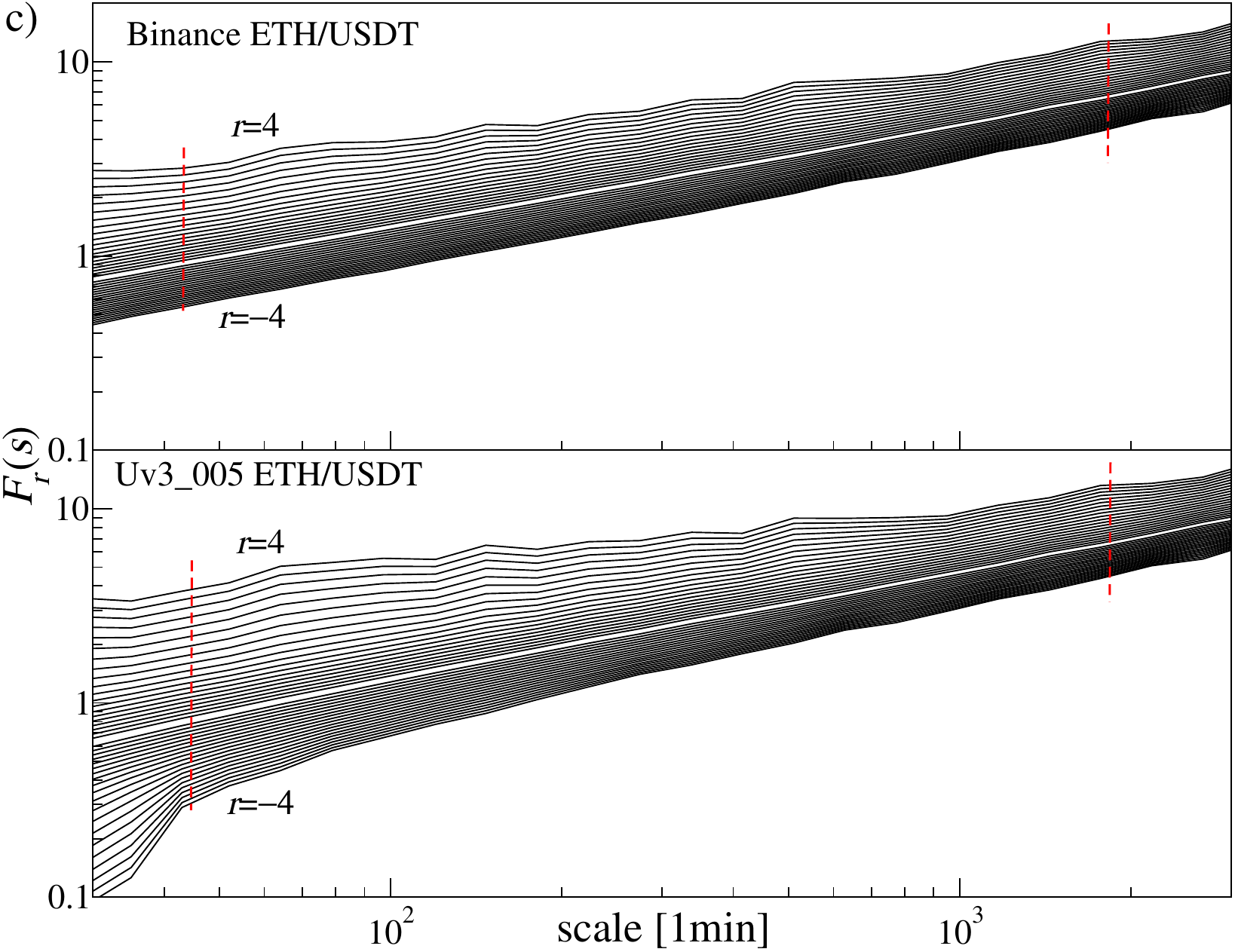}
\includegraphics[width=0.49\textwidth]{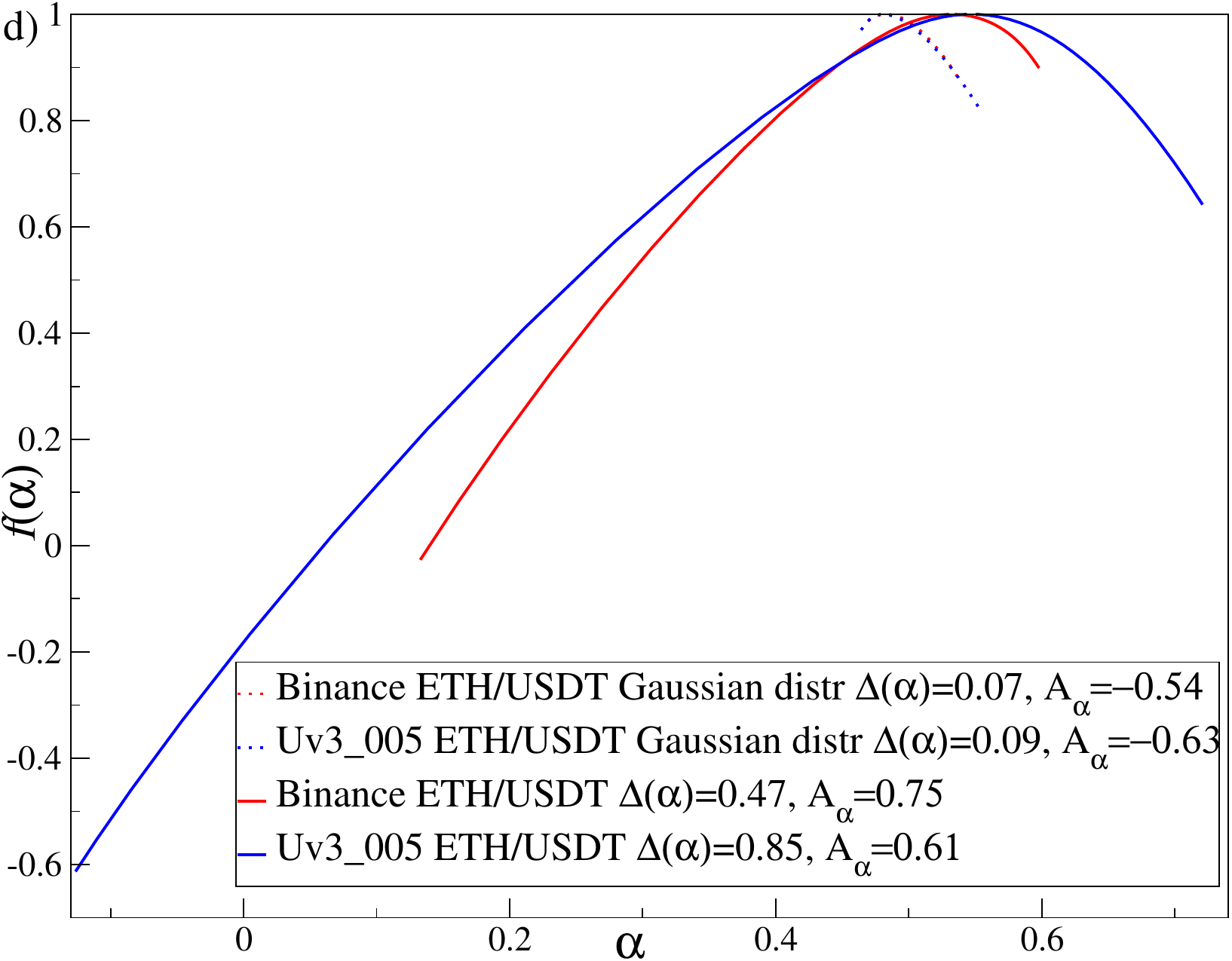}
\caption{Selected characteristics of ETH returns from Binance and Uniswap v3 with 0.05 provision covering the period from Jul 2024 to March 2025: a) - cumulative distribution function, b) - Pearson autocorrelation function, c) - fluctuation functions, d)  - multifractal spectra. Start of the scaling range, in which a power-law form of $F_r(s)$ is observed for a range of values of $q$, is denoted by a vertical red dashed line in c).}
\label{fig::Uniswap}
\end{figure}

As an extension of those studies, here we present a summary of the multifractal characteristics of the return time series for ETH on both markets (CEX from Binance and DEX from Uniswap v3 with 0.05 provision) in parallel. For this purpose, one-minute ETH returns covering the period from July 2024 to March 2025 are analyzed within the present extended methodology. The results are summarized in Fig.~\ref{fig::Uniswap} for the fluctuation PDFs (upper left), the autocorrelation function (upper right), the univariate fluctuation function (lower left) and the resulting multifractal spectra (lower right). It is worth noting that the ETH return PDFs have clearly thicker tails here than before, which refers even to those from the CEX market. A possible reason is that the DEX market, perhaps because it is less mature, fluctuates more rapidly, resulting in thicker tails. The CEX market, on the other hand, cannot be much different in this regard, as it would generate immediate arbitrage opportunities, meaning risk-free profits. Therefore, it is quite natural in this context to assume that the price dynamics on the CEX market is somewhat catching up and becoming more similar to that on the DEX market. Such a scenario would confirm the fact that the autocorrelation functions $A(\tau)$ appear almost identical in both markets. As a result, $F_r(s)$ for ETH from both markets show quite good and similar scaling, although the range of visible slopes (on a log-log scale — the lower left panel) for $r \geqslant 0$ is slightly larger on the DEX market than on the CEX market. This is reflected in the shapes of the multifractal spectra. Both are left-sided asymmetric, but the degree of asymmetry is clearly greater for the DEX market than for the CEX one. This fact is consistent with the presence of several larger events in the PDF corresponding to DEX with respect to CEX (see the upper left panel in Fig.~\ref{fig::Uniswap}). Finally, Gaussianization of these time series leads to an almost identical result with a spectrum that is still clearly multifractal but right-sided due to the suppression of the periods with large events.

\begin{figure}[ht!]
\centering
\includegraphics[width=0.49\textwidth]{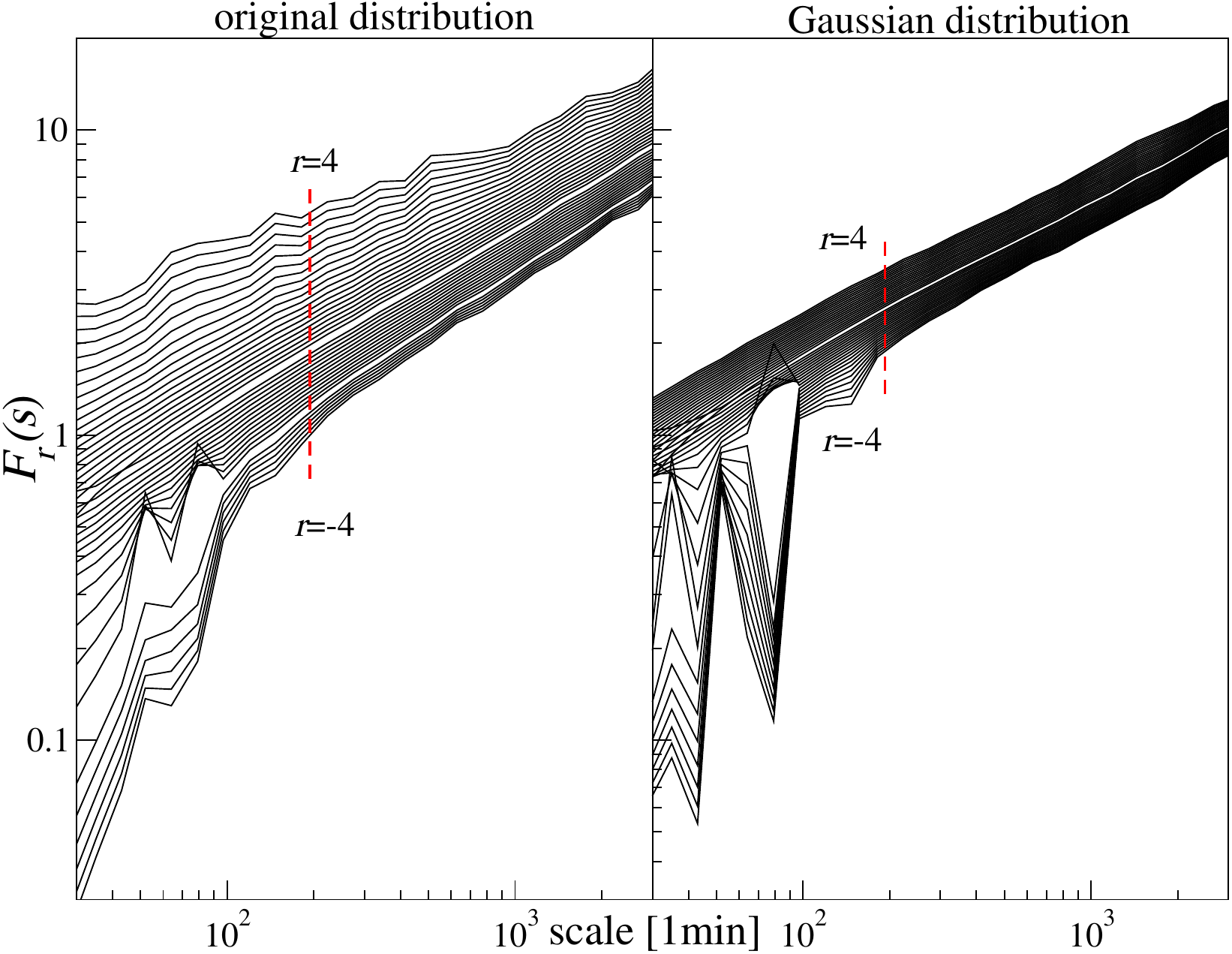}
\includegraphics[width=0.49\textwidth]{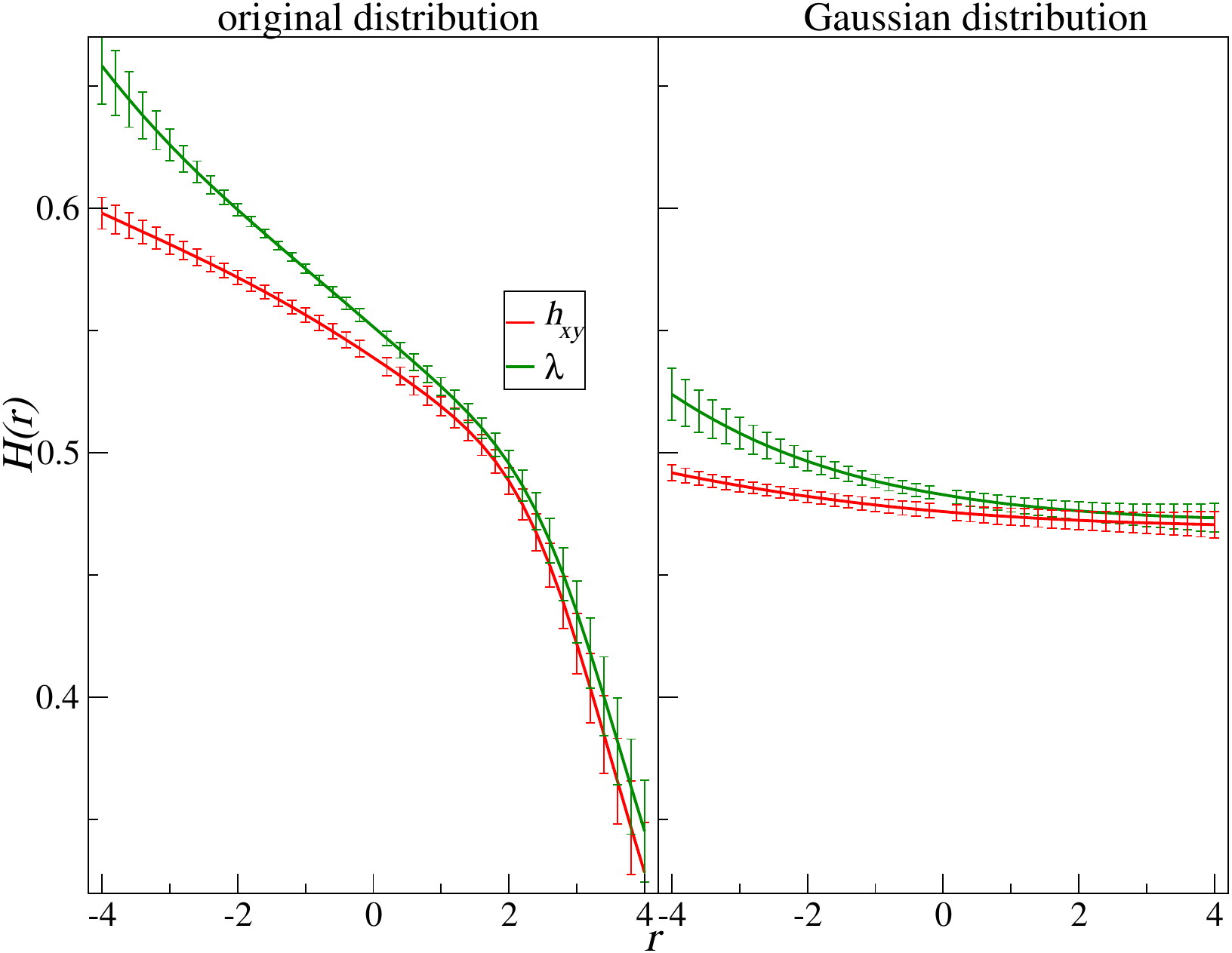}
\includegraphics[width=0.49\textwidth]{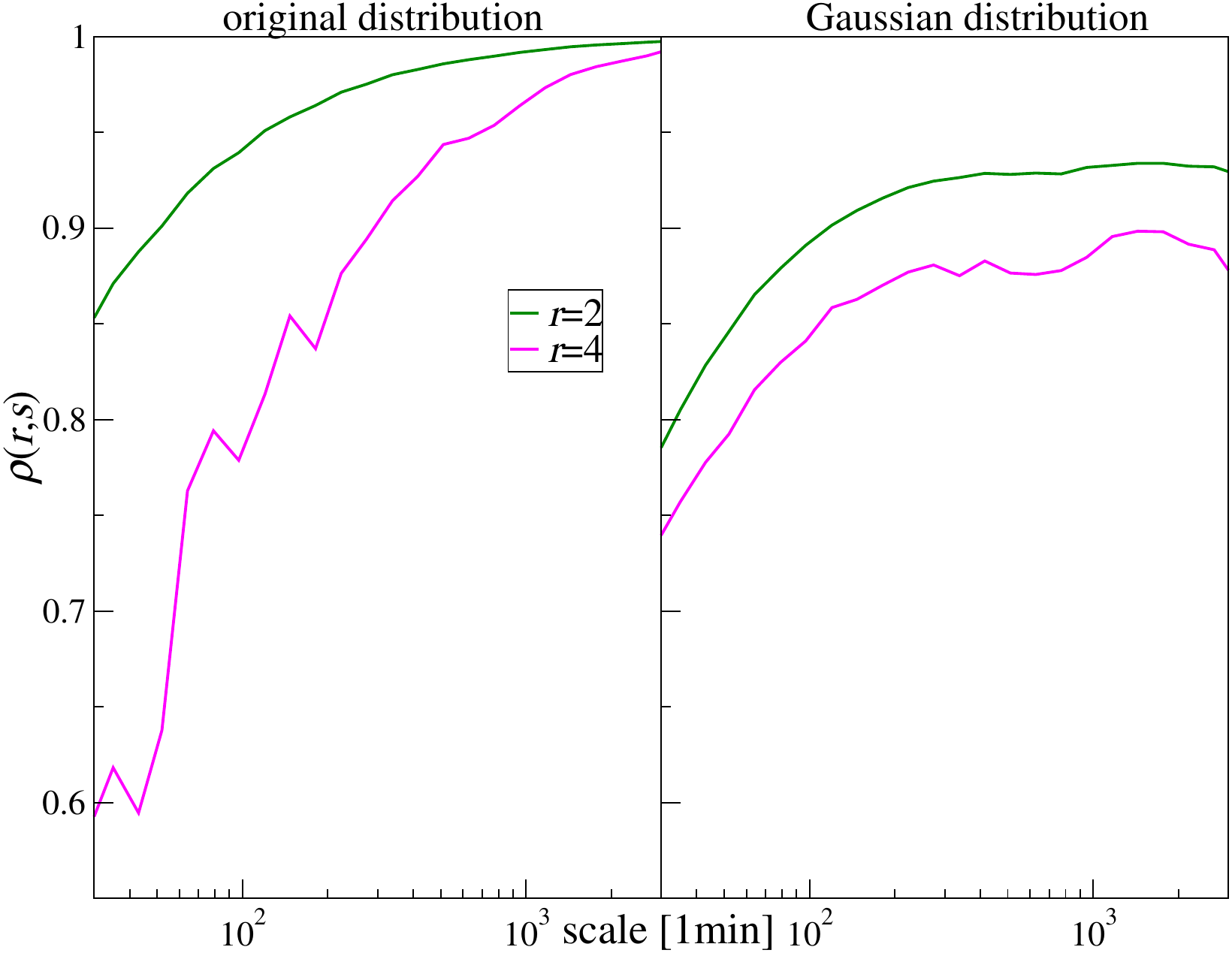}
\caption{Multiscale measures for time series of ETH returns collected from Binance and Uniswap v3 with 0.05 provision over the period Jul 2024 — March 2025: the bivariate fluctuation function $F_{\rm XY}(r,s)$ (top left), the generalized bivariate Hurst exponent $\lambda(r)$ and the average generalized univariate Hurst exponent $h_{\rm XY}(r)$ (top right), as well as the detrended cross-correlation coefficient $\rho_r(s)$ (bottom). The results for the original time series and their Gaussianized counterparts are presented in each case.}
\label{fig::Uniswapcc}
\end{figure}

Due to the fact that both the CEX and DEX time series show multifractal scaling, a natural further step is to consider their detrended cross-correlations as quantified by the bivariate fluctuation functions $F_{\rm XY}(r,s)$ and the coefficient $\rho_r(s)$. First, let us look at the former quantities, which are shown in the upper left panels of Fig.~\ref{fig::Uniswapcc}.  Unlike the data from the centralized market of Binance discussed in Sect.~\ref{sect::BTC.ETH.bivariate}, the fluctuation functions representing the original time series reveal scaling over the full range of the considered values of $r$, which is quite a remarkable result given the nature of each market is different. This may suggest that the cross-market arbitrage may play a crucial role in this case leading to a strong coupling between ETH returns on both markets. This is especially evident for $r \geqslant 0$ in the upper right panels of Fig.~\ref{fig::Uniswapcc}, which show the relation between $\lambda(r)$ and $h_{\rm XY}(r)$: both quantities converge to each other in this range of scales. What is even more convincing, $\rho_r(s)$ assumes values that are close to 1 already for a half-a-day-long time scale for $r=2$ (i.e., the periods of average fluctuations dominate) — it is a much stronger coupling than it was reported in Sect.~\ref{sect::BTC.ETH.bivariate} for the Binance data (see Fig.~\ref{fig::lambda_sred}). For the periods of large fluctuations $r=4$ this picture remains qualitatively analogous with the exception that the coupling becomes stronger for higher scales than for $r=2$. If the time series have been Gaussianized, the multifractal nature of data is still observed, although the variability of $\lambda(r)$ and $h_{\rm XY}(r)$  decrease, which indicates narrowing of the $f(\alpha)$ spectra in terms of $\Delta\alpha$. Also in this case there is no statistically significant difference between both these measures for positive values of $r$. However, for the Gaussianized time series, we observe evident weakening of the detrended cross-correlations, which assume values similar to those observed for the centralized market.


\subsection{NFT tokens}
\label{sect::NFT}

Another blockchain technology product of interest in light of the research conducted here is the non-fungible token (NFT) market. This market is a relatively new branch of blockchain-based finance that emerged in 2017, gaining mainstream attention with projects like CryptoKitties and marketplaces such as OpenSea. Unlike cryptocurrencies, which are interchangeable, each NFT is unique, representing ownership of digital or physical assets such as art, collectibles, and in-game items. The market experienced explosive growth during the COVID-19 pandemic, peaking in late 2021 with record-high trading volumes, before cooling down in subsequent years. Despite the NFT-market young age and volatility, research shows that NFT trading already displays statistical patterns similar to traditional financial markets, though with distinct features tied to low liquidity, token rarity, and speculative behavior~\cite{SzydloP-2024a,WatorekM-2024a}.

Trading on the NFT market differs from traditional financial markets in several important ways. Each token is unique, and its price often depends on rarity and individual traits rather than uniform value, making collections highly heterogeneous. Transactions are relatively infrequent, with long waiting times between trades, and activity tends to surge shortly after a collection’s launch before stabilizing. Instead of relying on last transaction prices, the NFT market often uses the floor price — the lowest ask price within a collection — as a key valuation metric, which makes it resemble an auction market. Recent study~\cite{SzydloP-2024a,WatorekM-2024a} shows that NFT floor prices, although less variable as compared to traditional asset prices, already carry complex internal dynamics similar to the other financial markets, with both non-linear correlations and heterogeneous behavior across scales.

\begin{figure}[ht!]
\centering
\includegraphics[width=0.49\textwidth]{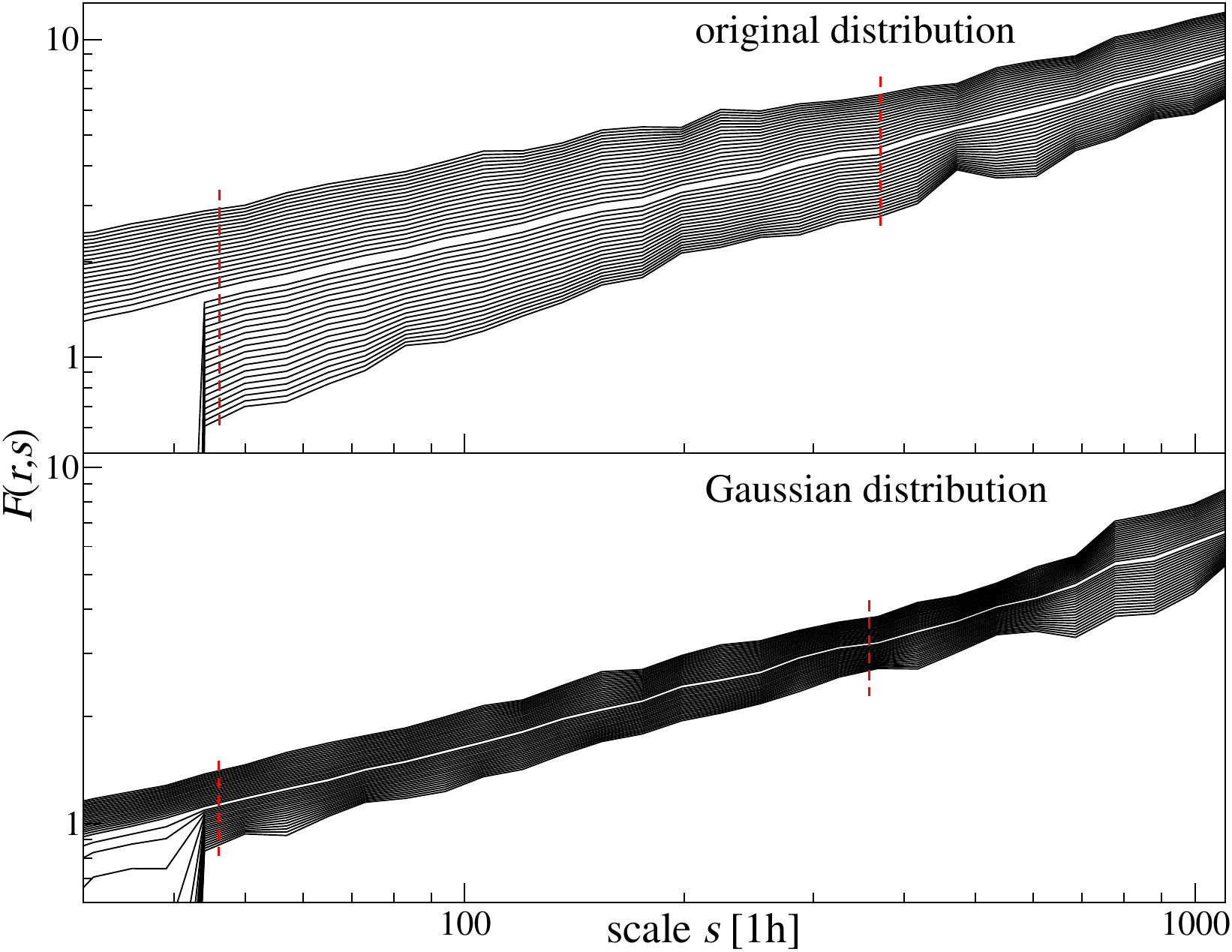}
\includegraphics[width=0.49\textwidth]{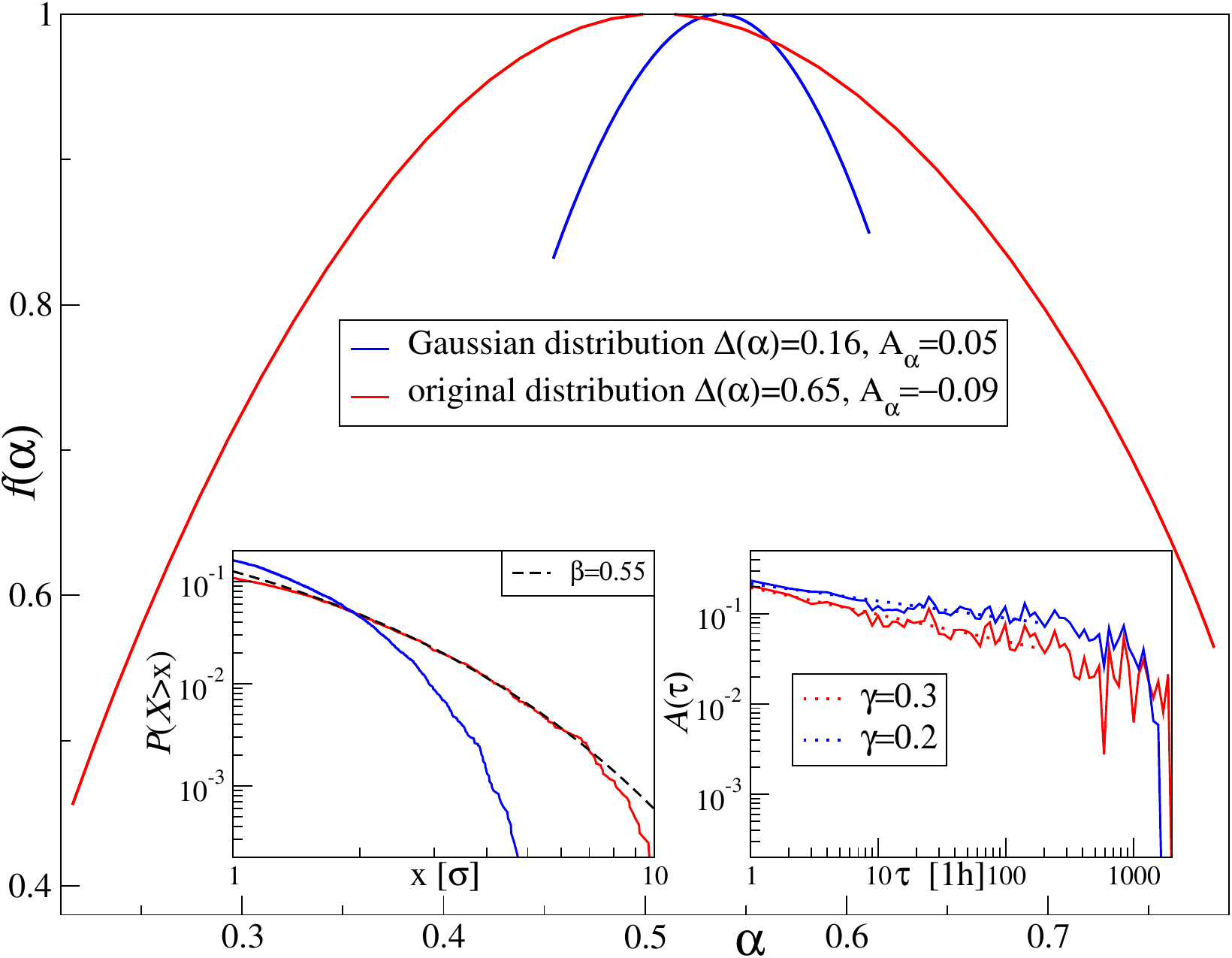}
\caption{The Famous Fox NFT collection floor price returns vs. their PDF-Gaussianized substitutes. Main panels: the univariate fluctuation functions $F(r,s)$ (left) and the multifractal spectra $f(\alpha)$ (right) calculated for the Famous Fox NFT collection floor price returns covering the period October 2021 -- September 2023. Insets: the PDFs (left) and the Pearson autocorrelation function (right) for the same data.}
\label{fig::NFT}
\end{figure}

As the last example, the Famous Fox Federation (FF) collection is subjected to a similar analysis as the ones above. This collection consists of 7,777 tokens on the Solana blockchain, each represented by digital images of stylized cartoon foxes with varying characteristics. The collection was developed by a group of Solana network users to build an ecosystem of blockchain-related tools and is associated with the FOXY token. In the statistical analysis~\cite{SzydloP-2024a,WatorekM-2024a}, the FF collection displayed heavy-tailed distributions in capitalization increments and transaction volumes, with some evidence of power-law scaling. Its trading activity also showed persistent correlations in capitalization and transaction volume, while floor price returns tended to be slightly antipersistent, reflecting the broader inefficiencies of the NFT market. In the current contribution, the FF price changes from the 2-year period between October 2021 and September 2023 were used and the results in a form analogous to the previous subsections are shown in Fig.~\ref{fig::NFT}. The insets on its right side show the PDFs and the autocorrelation function $A(\tau)$ for the original data and for their Gaussianized variants, respectively. The fluctuation functions (left panels) show multifractal scaling which translates into a wide $(\Delta \alpha = 0.65)$ multifractal spectrum (right panel, main) and substantially less asymmetric than for the time series analyzed in previous sections. It is interesting to note that the maximum of $f(\alpha)$ occurs slightly below $\alpha = 0.5$, which is an alternative manifestation of anti-persistence in NFT (here FF) returns. Gaussianization significantly reduces the $f(\alpha)$ width $(\Delta \alpha=0.16)$ but leaves it clearly in the multifractal region.


\section{Discussion and conclusions}
\label{sect::discussion}

Our study examined the multifractal properties of digital asset markets, focusing on Bitcoin, Ethereum, decentralized exchanges, and NFT tokens, by using the multifractal detrended cross-correlation analysis (MFCCA), the multifractal detrended fluctuation analysis (MFDFA) as its special case, and the related techniques. The results show convincingly that temporal correlations — particularly the long-range memory — constitute the fundamental source of multifractality in cryptocurrency price dynamics. While heavy-tailed return distributions play a complementary role by broadening the multifractal spectrum, they cannot by themselves generate multifractal behavior in the absence of correlations. Eliminating temporal correlations by shuffling the original time series leads to the disappearance of multifractality regardless of the thickness of the tails, while the singularity spectra are simply reduced to points, which confirms previously known results~\cite{DrozdzS-2009a,ZhouWX-2012a,KwapienJ-2023a}. Importantly, the temporal length of the digital market time series studied here is long enough to allow convergence to such a result. This finding strengthens the argument that genuine multifractality in financial time series reflects the persistent structural organization rather than statistical artifacts.

Across the seven-year period analyzed in this work (2018–2024), both BTC and ETH exhibited strong multifractal scaling, with notable variation during episodes of heightened uncertainty such as the COVID-19 pandemic. Their cross-correlations also displayed multifractal features, particularly for periods of large fluctuations, underlining the interconnectedness of the major cryptoassets. The comparative study of centralized (Binance) and decentralized (Uniswap) exchanges revealed that while both markets show multifractality, decentralized markets exhibit stronger left-skewed spectra driven by periods of large fluctuations, consistent with their relative immaturity and lower liquidity. Extending the analysis to the NFT sector represented here by the Famous Fox collection highlighted that even young, heterogeneous markets can already show developed multifractal scaling, even more symmetric as compared to cryptocurrencies. This may indicate that the uncorrelated noise component in NFTs is smaller than in cryptocurrencies.

Taken together, these results reinforce the importance of applying multifractal frameworks to digital markets. They offer practical insights for volatility forecasting, systemic risk monitoring, and understanding the maturation of blockchain-based financial ecosystems. For example, the strong correlations between BTC and ETH on various time scales can be used in optimal portfolio construction~\cite{zhao2018}. The decay of the autocorrelation function, which indicates the average size of the volatility cluster~\cite{DrozdzS-2009a}, may also be used in risk management. The consistency of the log-return tail distributions with the inverse cubic power-law allows for the selection of appropriate distributions in various risk mitigation methods, such as Value-at-Risk. Moreover, the greater hierarchical dependence of correlations at the level of larger fluctuations compared to smaller ones, documented by left-sided asymmetry of the multifractal spectra, may potentially allow for extending risk management strategies to include fluctuation-specific correlation patterns. The direct link between multifractality and risk management is an interesting subject for further studies.

On a more general level, the disentangling methodology used here — separating distributional effects from temporal correlations — provides a robust foundation for future research, enabling a clearer distinction between genuine complexity and spurious effects. As digital finance continues to expand, multifractal analysis can serve as a valuable tool for interpreting its evolving dynamics, informing both academic inquiry and market practice.

\vspace{6pt} 

\authorcontributions{Conceptualization, S.D., R.K, J.K. and M.W.; methodology, S.D., R.K., J.K. and M.W.; software, R.K. and M.W.; validation, S.D., J.K. and M.W.; formal analysis, R.K. and M.W.; investigation, S.D., J.K. and M.W.; resources, M.W.; data curation, M.W.; writing—original draft preparation, S.D. ; writing—review and editing, J.K. and M.W.; visualization, M.W.; supervision, S.D. All authors have read and agreed to the published version of the manuscript.}

\funding{This research received no external funding.}

\dataavailability{Data available freely from Binance~\cite{binance}.} 

\conflictsofinterest{\hl{ }The authors declare no conflicts of interest.}

\begin{adjustwidth}{-\extralength}{0cm}

\reftitle{References}

\bibliography{refs}

\end{adjustwidth}

\end{document}